\documentclass[twocolums]{aa}  

\usepackage{graphicx}
\usepackage{txfonts}
\usepackage{lipsum}
\usepackage{subcaption}         
\usepackage{lscape}             
\usepackage{placeins}           
\usepackage{natbib}

\usepackage{booktabs}   

\usepackage{url}
\usepackage{xcolor}

\begin{document}

\title{Comparison of Hybrid and Standard Reconstruction Methods on IXPE Astrophysical Data}

%
%
%

   \author{R. Ferrazzoli\inst{1}\corrauth{riccardo.ferrazzoli@inaf.it}        
        \and N. Cibrario\inst{2}
        \and S. Silvestri\inst{3}
        \and M. Negro\inst{4}
        \and P. Soffitta\inst{1}
        \and S. Chakraborty \inst{5}
        \and S. Pagliarella\inst{1} \inst{6} \inst{7}
        \and S. R. Ehlert \inst{8}
        }

   \institute{INAF Istituto di Astrofisica e Planetologia Spaziali, Via del Fosso del Cavaliere 100, 00133 Roma, Italy
   \and Istituto Nazionale di Fisica Nucleare, Sezione di Torino, Via Pietro Giuria 1, 10125 Torino, Italy 
   \and Istituto Nazionale di Fisica Nucleare, Sezione di Pisa, Largo B. Pontecorvo 3, 56127 Pisa, Italy
   \and Department of Physics and Astronomy, Louisiana State University, Baton Rouge, LA 70803, USA 
   \and Science and Technology Institute, Universities Space and Research Association, Huntsville, AL 35805, USA   
   \and Department of Physics, Sapienza University of Rome, Piazzale Aldo Moro 5, 00185 Rome, Italy
   \and Department of Physics, Tor Vergata University of Rome, Via della Ricerca Scientifica 1, 00133 Rome
   \and NASA Marshall Space Flight Center, 320 Sparkman Drive, Huntsville, AL 35812, USA   
   }

   \date{}

 
  \abstract
   {The hybrid reconstruction algorithm developed by \citet{2023Cibrario} combines a convolutional neural network (CNN) with standard moment analysis to improve the reconstruction of photoelectron tracks in the Gas Pixel Detectors (GPDs) on board the Imaging X-ray Polarimetry Explorer (IXPE). 
   The method has been validated on laboratory calibration data in \citet{2025Cibrario}, but its performance on astrophysical sources with known polarimetric properties has not yet been systematically characterized.
   }
   {We present the first comparison of hybrid and standard reconstruction methods on IXPE data of astrophysical point sources, with the aim of quantifying the improvement in polarimetric sensitivity as a function of energy. 
   } 
   {We analyze two IXPE observations selected to probe complementary regimes: Her X-1, a bright and highly polarized accreting X-ray pulsar used as a quantitative benchmark, and the transient neutron star low-mass X-ray binary XTE J1701-461 in its soft state, a bright source with no detectable polarization used to measure residual systematic effects. 
   For each source, we compare the normalized Stokes parameters, the degree of polarization (PD) and angle (PA), and the detection significance derived from the level 2 event files produced by both pipelines. 
   }
   {For Her X-1, the hybrid method yields a consistently higher detection significance at all energies, with a statistically significant improvement above $\sim$2.83 keV. 
   The significance ratio $R = \sigma_\mathrm{hyb} /\sigma_\mathrm{std}$ deviates from unity by up to $9.3\sigma$ in the 2.83-4.00 keV bin, rising to $R = 1.18 \pm 0.03$ at 5.66-8.00 keV at 6.2$\sigma$.
   For XTE J1701-461, both methods yield polarization degrees consistent with zero and significance ratios consistent with unity. 
   Through a Fourier analysis we identify some modes ($m = 3$, $m = 6$, $m = 12$, and $m = 15$) with power excesses in the hybrid angle distributions at low energies, significant at $>7\sigma$ relative to the standard method, while at 5-8 keV all modes are consistent with the expected noise level for both methods: these modes are amplified by CNN but do not propagate to the Stokes parameters.
   }
   {The hybrid method provides a measurable and statistically significant improvement in polarimetric sensitivity for bright point sources that is fully consistent with the predictions of \citet{2025Cibrario} and equivalent to a $\sim20\%$ increase in effective area without the additional cost of observation time. 
   The hexagonal-grid modes are a geometric artifact of the GPD pixel structure and have no effect on polarimetric results since the additional modes are orthogonal to the polarization modes.
   The improvement is robustly detected for the bright source examined here ($\sigma_\mathrm{std}\gtrsim9$). 
   For the unpolarized source in our sample ($\sigma_\mathrm{std}\sim1$-$2$) the central significance increase is consistent with the same trend, but is not individually resolved at the available statistics. 
   These findings confirm the readiness of the hybrid method for systematic application to IXPE astrophysical data.
   }

   \maketitle

\nolinenumbers

\section{Introduction}
\label{sec:introduction}
The Imaging X-ray Polarimetry Explorer \citep[IXPE;][]{2022Weisskopf,2021Soffitta}, launched in December 2021, has opened a new observational window on the polarized X-ray sky, allowing spatially resolved \textit{spectro-polarimetric} measurements in the 2-8 keV band for the first time. 
The instrument consists of three identical detector units, each comprising a mirror module and a Gas Pixel Detector \citep[GPD;][]{2001Costa, 2006Bellazzini, 2021Baldini}, in which the direction of linear polarization of incoming X-rays is encoded in the azimuthal distribution of the photoelectron emission angle $\phi$. 
The standard event reconstruction method, known as moment analysis \citep{2003Bellazzini}, estimates both the photoelectron emission angle and the photon impact point (IP) from the morphological properties of the ionization track left by the photoelectron in the detector gas.

\citet{2023Cibrario} proposed a hybrid analytic-machine learning (ML) approach in which a convolutional neural network (CNN) based on the DenseNet-121 architecture \citep{2016Huang}, incorporating hexagonal convolutional layers  \citep{2019Steppa}, is trained to predict the photon IP from the track image. 
By substituting this predicted IP from CNN into the standard moment analysis, the hybrid method improves the modulation factor by $\sim$1\% at 3 keV and $\sim$6\% at 6 keV on Monte Carlo simulations. 
A further improvement was achieved by introducing an artificial image-sharpening step prior to CNN inference.
\citet{2025Cibrario} subsequently validated the algorithm on laboratory calibration data from the IXPE pre-launch campaign \citep{2022Muleri,2022DiMarco}, confirming the expected improvement in modulation factor and demonstrating a significant mitigation of the \emph{polarization leakage} effect \citep{2023Bucciantini_a} for extended sources such as the Pulsar Wind Nebula G21.5-0.9 \citep{2025DiLalla}.

A critical step in characterizing any new reconstruction method for IXPE is its application to astrophysical sources with well-known polarimetric properties. 
Laboratory validation of \citet{2025Cibrario} used unpolarized and polarized monochromatic spatially uniform beams under controlled conditions. 

However, astrophysical observations involve broad-band spectra, source-dependent spatial distributions, and variable background levels that may affect the relative performance of the two methods in ways not captured by laboratory tests. 
In particular, no systematic comparison with bright point sources (for which the polarimetric sensitivity improvement is most directly measurable) has been done to date.

In this work, we present the first comparison of hybrid and standard reconstruction methods on IXPE astrophysical data, using two point sources that probe complementary polarimetric regimes:
\begin{itemize}
  \item \textbf{Her X-1} is a bright and highly polarized accreting X-ray pulsar with a well-characterized polarization degree of $\sim$8-10\% across the IXPE band \citep[]{2022Doroshenko,2023Garg,2024Heyl,2024Zhao}. 
  Its high count rate, relatively hard spectrum, and large polarization signal make it an ideal quantitative benchmark for comparing the sensitivity of the two reconstruction methods as a function of energy.
  \item \textbf{XTE J1701-461} is a transient neutron star low-mass X-ray binary. Its soft state is bright and its 2–8 keV polarimetric signal could not be constrained, with a 99\% confidence level upper limit of 1.5\% \citep[OBSID 01250701;][]{2023Cocchi}, providing a measurement of residual systematic effects under conditions where the true polarimetric signal is (close to) zero. 
  This is analogous to the analysis of unpolarized beam measurements performed in \citet{2025Cibrario} and allows us to characterize the hexagonal-pixel systematic modes ($m = 6$, $m = 12$) that appear in the photoelectron angle distribution of the hybrid method at low energies.
  We stress that the classification of a source as "unpolarized" is necessarily relative to the statistical precision of the measurement, not an absolute property: as the statistical uncertainty on PD is reduced (e.g.\ with longer exposures, brighter sources, or more sensitive reconstruction methods), an observationally unpolarized source can in principle reveal either a residual instrumental systematic or a true, low-level polarization. 
\end{itemize}
 
The paper is organized as follows. 
Section \ref{sec:methods} describes the observations, data reduction, and comparison methods. 
Section \ref{sec:results} presents the polarimetric results and the Fourier analysis of the angle distributions. 
We discuss the implications in Section \ref{sec:discussion}.

\section{Observations and methods}
\label{sec:methods}
 
\subsection{Observations and data reduction}
\label{subsec:data}
Both sources were observed by IXPE with all three detector units (DUs).
Table\ref{tab:observations} summarizes the observations used in this work.
\begin{table}[htbp]
  \centering
  \caption{IXPE observations analyzed in this work.}
  \label{tab:observations}
  \begin{tabular}{lccc}
    \hline
    Source            & OBSID    & Exposure (ks) & Date       \\
    \hline
    Her X-1           & 02004001  & 244.5           & 2023-02-03        \\
    XTE J1701-461     & 01250701  & 46.4           & 2022-10-08        \\
    \hline
  \end{tabular}
\end{table}
 %
The required software ecosystem as well as the processing pipeline that we used is implemented in a publicly available dedicated container \citep{container}\footnote{Available at \url{https://doi.org/10.5281/zenodo.21395803}} that can run the full analysis from level 1 to level 2 and also includes an ixpeobssim version with the hybrid response functions.
To ensure a consistent comparison between the standard and hybrid reconstructions, we did not use the level 2 files distributed by HEASARC. Instead, we reprocessed the observations starting from the original level 1 event files, since the standard IXPE level 1 to level 2 processing pipeline has undergone updates throughout the mission lifetime. 
We obtained the \emph{standard} files by processing the level 1 files distributed by HEASARC with the publicly available tool \emph{ixpepipe}\footnote{Available at \url{https://heasarc.gsfc.nasa.gov/docs/ixpe/analysis/contributed/ixpepipe.html}}. 
For \emph{hybrid} reconstruction, we first generated modified level 1 files by replacing the IP determined by the standard moment analysis with the IP predicted by the convolutional neural network. 
The photoelectron emission angle was then recomputed by running the moment analysis employing the new CNN-predicted IP. 
The resulting \emph{hybrid} level 1 files were subsequently processed with \emph{ixpepipe} following the same procedure adopted for the standard reconstruction. 
The hybrid files use dedicated spurious modulation correction maps and modulation factor curves generated by applying the hybrid reconstruction pipeline to the IXPE pre-launch laboratory calibration data \citep{2022Muleri,2022DiMarco}, so that they account for the different systematic response of the CNN-predicted impact point relative to the standard moment-analysis estimate. 
The modulation factor curves are additionally corrected for the GPD gas pressure at the epoch of each observation, following the same procedure employed for the standard analysis.
We perform polarimetric analysis of the level 2 event files with the \texttt{ixpeobssim} package \citep[v32.0.1;][]{2022Baldini} using the \texttt{PCUBE} algorithm. 
We used for both observations the instrument response functions \emph{ixpe:obssim20230101:v013}, in the case of the hybrid files, we used the improved modulation factor from the hybrid method.

We extract source events from a circular region centered on each source with radius 2.5'.
Due to the brightness of the sources considered, no background subtraction or rejection is needed, as prescribed by \citet{2023DiMarco}.
We perform the analysis in four logarithmically spaced energy bins covering the 2.00-8.00 keV range. 
We describe the analysis of Level 1 event files, used for the Fourier study of the angle distribution, in Section \ref{subsec:fft_method}.

\subsection{Significance ratio and its uncertainty}
\label{subsec:significance}
 
The significance of measuring a certain polarization degree PD is $\rm \sigma = PD/PD_{err}$, where PD and its uncertainty $\rm PD_{err}$ are derived from the normalized Stokes parameters following \citet{2015Kislat}.
To quantify the improvement \textit{due to} the hybrid method, we define the significance ratio $R$:
\begin{equation}
  R = \frac{\sigma_\mathrm{hyb}}{\sigma_\mathrm{std}} = \frac{\mathrm{PD}_{\rm hyb}}{\mathrm{PD}_{\rm std}}
\frac{\mathrm{PD}_{{\rm err,std}}}{\mathrm{PD}_{{\rm err,hyb}}}
\equiv R_{\rm PD}R_{\rm err},
  \label{eq:R}
\end{equation}
where $\sigma_\mathrm{hyb}$ and $\sigma_\mathrm{std}$ are the significance of the hybrid and standard measurement, respectively.
A value $R > 1$ indicates that the hybrid method achieves a higher significance for detection in that energy bin.
The factor $R_{\rm err}$ quantifies the improvement in the statistical precision, whereas $R_{\rm PD}$ depends on the particular realization of the measured polarization degree. 
Therefore, $R$ can be affected by statistical fluctuations in the central PD values even when the two reconstruction methods have identical expectation values.\\
The uncertainty in $R$ requires careful treatment. 
Since $\sigma_\mathrm{hyb}$ and $\sigma_\mathrm{std}$ are derived from the same photons processed through two different pipelines, they are not independent. 
The propagation of standard error for a ratio of independent quantities would substantially overestimate $\sigma(R)$. 
The correct formula, accounting for the covariance between the two estimators, is then:
\begin{equation}
  R_{err} = \frac{\sqrt{{\sigma_{hyb, err}}^2
              + R^2\,{\sigma_{std,err}}^2
              - 2R\,\mathrm{Cov}(\sigma_{hyb}, \sigma_{std})}}
              {\sigma_{std}}\,.
  \label{eq:R_err_full}
\end{equation}
With $\mathrm{Cov}(\sigma_{hyb}, \sigma_{std}) = \rho\, \sigma_{\sigma_{hyb}}\,\sigma_{\sigma_{std}}$ and using the result that in the high-statistic limit (number of effective photons $N_\mathrm{eff} \gg \sigma^2 / 2\mu^2$, verified for all our energy bins \footnote{e.g. in the lowest statistics energy bin, 5.66-8.00 keV of Her X-1: $N_\mathrm{eff}= 53807$ is much larger than $\sigma^2 / 2\mu^2= 88$}), the uncertainty on the significance  
\begin{equation}
    \sigma_{err} = \sqrt{1 + \frac{PD_\mathrm{err}^2}{2 N_\mathrm{eff} \mu^2}}
\end{equation}
satisfies $\sigma_{std,err} \sim \sigma_{hyb,err} \approx 1$ \citep{2006Vaillancourt}, and Eq. \eqref{eq:R_err_full} simplifies to:
\begin{equation}
  R_{err} = \frac{\sqrt{1 + R^2 - 2R\rho}}{\sigma_{std}}\,.
  \label{eq:R_err}
\end{equation}
The correlation coefficient $\rho$ is estimated directly from the normalized Stokes parameters $q$ and $u$:
\begin{equation}
  \rho \approx
    \frac{q_\mathrm{hyb}\,q_\mathrm{std} + u_\mathrm{hyb}\,u_\mathrm{std}}
         {\mathrm{PD}_\mathrm{hyb} \cdot \mathrm{PD}_\mathrm{std}}\,.
  \label{eq:rho}
\end{equation}
For both sources, we find $\rho \approx 0.999$ in all energy bins, as expected since the two pipelines differ only in the IP prediction and share the same underlying photon events. 
Neglecting this correlation would overestimate $R_\mathrm{err}$ by a factor of $\sim 1/\sqrt{1 - \rho^2} \approx 20$.

The significance ratio $R$ also allows us to express the improvement in terms of an equivalent increase in the mirror effective area or detector efficiency. 
Since $\sigma \propto \sqrt{A_\mathrm{eff}}$, under the hypothesis that the PD is the same under the two methods, the fractional increase in effective area that would be needed for the standard method to match the hybrid significance is:
\begin{equation}
  \frac{\Delta A_\mathrm{eff}}{A_\mathrm{eff}} = R^2 - 1\,,
  \label{eq:DAeff}
\end{equation}
with uncertainty:
\begin{equation}
  \left(\frac{\Delta A_\mathrm{eff}}{A_\mathrm{eff}}\right)_\mathrm{err}
  = 2\,R\,R_\mathrm{err}\,.
  \label{eq:DAeff_err}
\end{equation}
This quantity provides us with a measure of the gain delivered by the hybrid method that is directly comparable across different energy bands and sources.
Because $R$ contains both the change in statistical precision and the realization-dependent change in the measured PD, this quantity should be interpreted as an observed effective-area-equivalent gain rather than as a direct measurement of an instrumental effective-area increase.
A complementary figure of merit that depends only on the measurement precision is the minimum detectable polarizations \citep[MDP,][]{2010Weisskopf}:
\begin{equation}
MDP = \frac{4.29}{\mu \sqrt{C_s}} ,
  \label{eq:MDP}
\end{equation}
where $\rm \mu$ is the modulation factor and $\rm C_s$ the source counts.
This quantity, valid for the current case of bright sources, is independent of the particular PD realization and captures both the reduction in statistical uncertainty and the improvement in the modulation factor.
We can therfore explicitly define an MDP ratio,
\begin{equation}
R_{\rm MDP}=
\frac{\mathrm{MDP}_{\rm std}}{\mathrm{MDP}_{\rm hyb}}.
\end{equation}
We report both quantities in Table \ref{tab:herx1_sig} for Her X-1 and Table \ref{tab:xte_sig} for XTE J1701-461.
We also note that $\sigma = \mathrm{PD}/\mathrm{PD}_{\rm err}$ has a simple Gaussian interpretation only when $\sigma \gg 1$.
For low-significances it should be understood as a detection-test statistic and all PD values for that source are formally upper limits (see for example Section \ref{subsec:xte}).

\subsection{Fourier analysis of the photoelectron angle distribution}
\label{subsec:fft_method}
To characterize systematic effects in the photoelectron angle distribution at the track level, we directly analyze the IXPE level 1 event files. 
For the unpolarized source XTE J1701-461, in each logarithmically binned energy band, we extract the reconstructed emission angle $\phi$ from the \texttt{DETPHI2} column, applying an energy selection via event time-matching with the corresponding level 2 files.
We combine the three detector units after applying nominal rotations of $0\degr$, $120\degr$, and $240\degr$ to account for the clocking of each DU in the focal plane.
We compute the power spectrum of the angle distribution $N(\phi)$ via a discrete Fourier transform after subtracting the mean count rate (i.e. the $m=0$ component), so that the power spectrum reflects only the periodic modulations of the distribution.
The power at mode $m$ corresponds to a periodicity of $360\degr/m$ in the angle distribution. 
Mode $m = 2$ carries the polarization signal, while modes $m = 6$ and $m = 12$ are associated with the hexagonal symmetry of the GPD pixel grid \citep[Appendix B]{2025Cibrario}.

\section{Results}
\label{sec:results}
 
\subsection{Polarimetric results: Her X-1}
\label{subsec:herx1}
\begin{figure*}[htbp!]
  \centering
  \includegraphics[width=\linewidth]{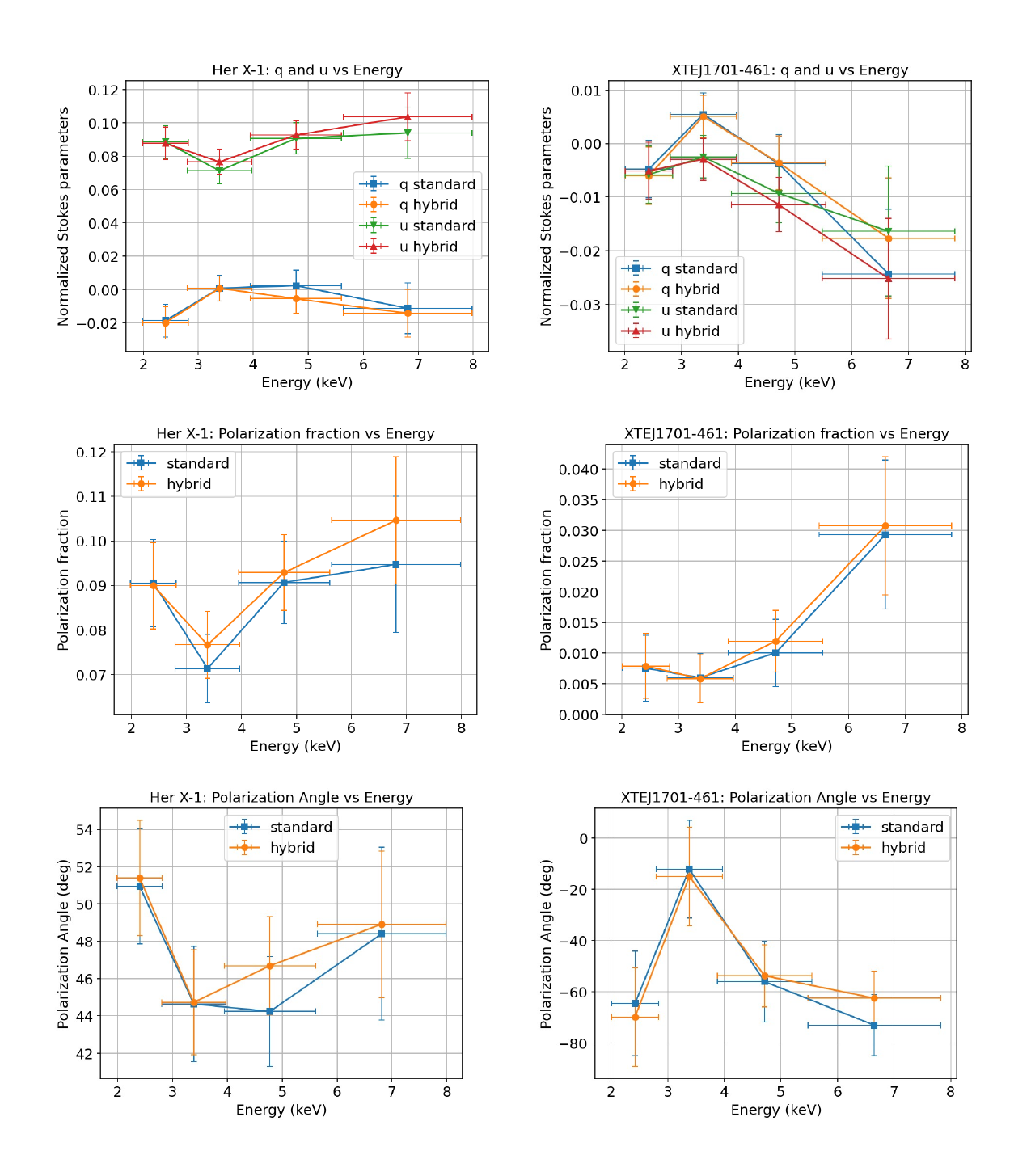}
  \caption{Normalized Stokes parameters $q$ and $u$ (\emph{top row}), polarization degree PD (\emph{middle row}), and polarization angle PA (\emph{bottom row}) as a function of energy for Her X-1 (\emph{left column}) and XTE J1701-461 (\emph{right column}).
  For Her X-1, both methods yield consistent results at all energies, with the hybrid method producing smaller uncertainties above $\sim$3 keV. 
  For XTE J1701-461, all measurements are consistent with zero polarization. 
  We highlight that the displayed PD values and PA angles are not formally significant and are shown solely to demonstrate the mutual consistency of the two pipelines.
  }
  \label{fig:pol_properties}
\end{figure*}

\begin{figure}[htbp]
  \centering
  \includegraphics[width=1\linewidth]{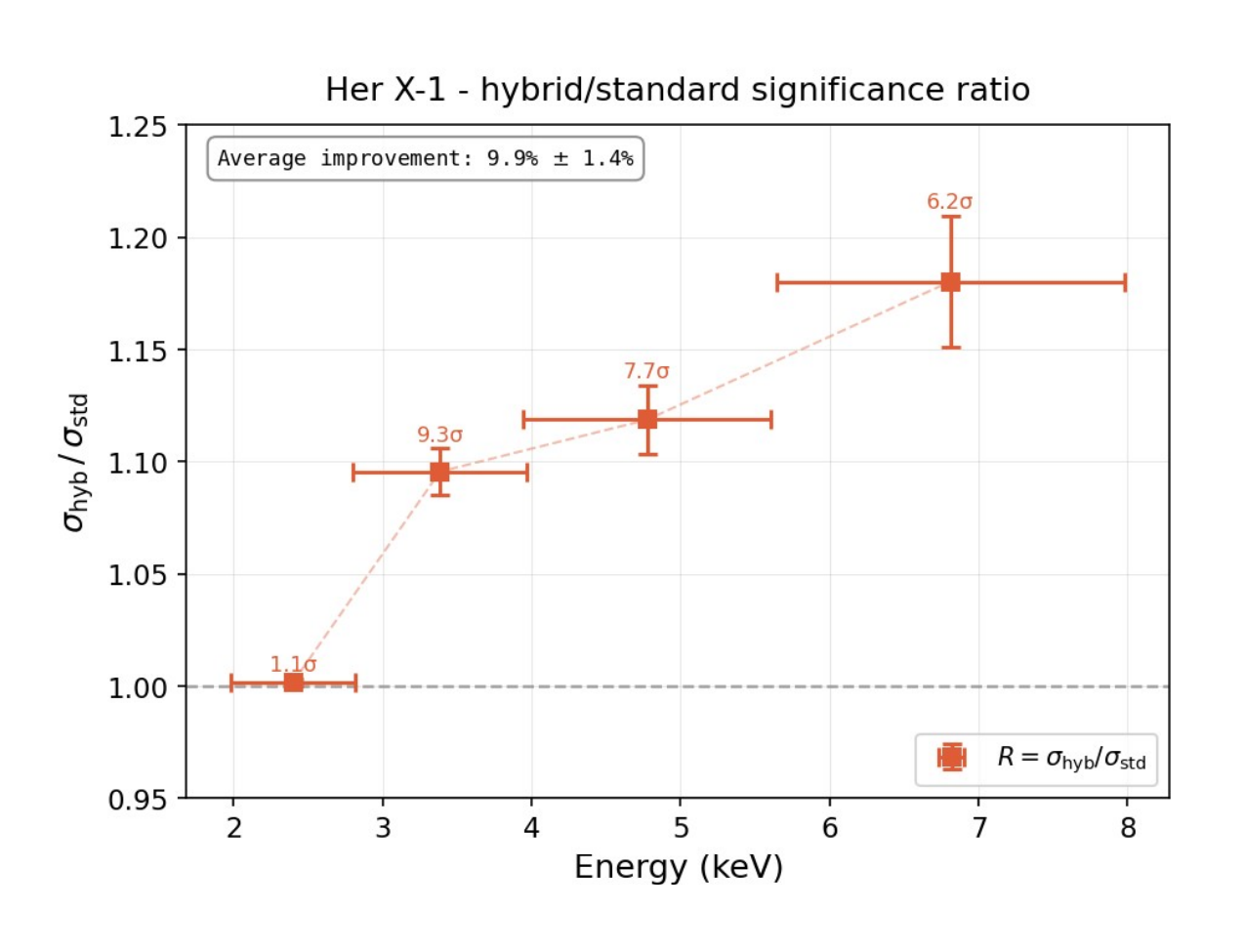}
  \caption{%
    Ratio of detection significances $R = \sigma_\mathrm{hyb}/\sigma_\mathrm{std}$ for Her X-1 as a function of energy. 
    Error bars are computed from Eq.~\eqref{eq:R_err}.
    The dashed horizontal line marks $R = 1$ (equal performance).
    The average improvement over the full 2-8 keV energy band is of $9.9\%\pm1.4\%$ }
  \label{fig:significance_ratio_HerX1}
\end{figure}

We detected polarization from Her X-1 with high significance in all four energy bins with both reconstruction methods. 
Fig.\ref{fig:pol_properties} (left column) shows the normalized Stokes parameters $q$ and $u$, PD, and PA as a function of energy for both methods.
Both methods yield consistent polarimetric observables values within their respective uncertainties at all energies, with no statistically significant differential systematic offset between the two pipelines. 
The hybrid method produces noticeably smaller error bars above $\sim$3 keV, reflecting the improved modulation factor.
We report the normalized Stokes parameters $q$ and $u$, PD, and PA in Table \ref{tab:herx1_pol} .
\begin{table*}[htbp]
  \centering
  \caption{%
    Polarimetric parameters ($q$, $u$, PD, and PA) for Her X-1 as a function of energy, derived with the standard moment analysis (Std) and the hybrid method (Hyb). 
    All quantities are derived with \texttt{PCUBE}.
  }
  \label{tab:herx1_pol}
  \setlength{\tabcolsep}{4pt}
  \begin{tabular}{l cc cc cc cc}
    \toprule
    & \multicolumn{2}{c}{$q$ (\%)}
    & \multicolumn{2}{c}{$u$ (\%)}
    & \multicolumn{2}{c}{PD (\%)}
    & \multicolumn{2}{c}{PA (\degr)} \\
    \cmidrule(lr){2-3}\cmidrule(lr){4-5}
    \cmidrule(lr){6-7}\cmidrule(lr){8-9}
    $E$ (keV) & Std & Hyb & Std & Hyb & Std & Hyb & Std & Hyb \\
    \midrule
    2.00--2.83 & $-1.87{\pm}0.98$ & $-1.99{\pm}0.97$ & $8.86{\pm}0.98$ & $8.77{\pm}0.97$ & $9.05{\pm}0.98$ & $8.99{\pm}0.97$ & $51.0{\pm}3.1$ & $51.4{\pm}3.1$ \\
    2.83--4.00 & $0.09{\pm}0.77$ & $0.07{\pm}0.75$ & $7.13{\pm}0.77$ & $7.66{\pm}0.75$ & $7.13{\pm}0.77$ & $7.66{\pm}0.75$ & $44.6{\pm}3.1$ & $44.7{\pm}2.8$  \\
    4.00--5.66 & $0.24{\pm}0.93$ & $-0.55{\pm}0.85$ & $9.06{\pm}0.93$ & $9.27{\pm}0.85$ & $9.07{\pm}0.93$ & $9.29{\pm}0.85$ & $44.2{\pm}2.9$ & $46.7{\pm}2.6$ \\
    5.66--8.00 & $-1.12{\pm}1.53$ & $-1.43{\pm}1.43$ & $9.40{\pm}1.53$ & $10.37{\pm}1.43$ & $9.47{\pm}1.53$ & $10.46{\pm}1.43$ & $48.4{\pm}4.6$ & $48.9{\pm}3.9$ \\
    \bottomrule
  \end{tabular}
\end{table*}
We report the significance of detection and its comparison between the two methods in Table \ref{tab:herx1_sig}. 
Fig. \ref{fig:significance_ratio_HerX1} shows the significance ratio $R = \sigma_\mathrm{hyb}/\sigma_\mathrm{std}$ as a function of energy.
%

\begin{table*}[htbp]
  \centering
  \caption{%
    Detection significance comparison for Her X-1.
    Energy bins are logarithmically spaced in the 2-8 keV range.
    MDP$_\mathrm{std}$ and MDP$_\mathrm{hyb}$ are the minimum detectable polarizations and R$_\mathrm{MDP}$ their ratio};
    $R = \sigma_\mathrm{hyb}/\sigma_\mathrm{std}$; $R_\mathrm{err}$ is computed
    from Eq. \ref{eq:R_err};
    $(R-1)/R_\mathrm{err}$ is the deviation from parity;
    $\Delta A_\mathrm{eff}/A_\mathrm{eff}$ is the equivalent
    fractional increase in effective area (Eq. \ref{eq:DAeff}).
    In all energy bins $R>1$, with high significance above 2.83 keV.
  \label{tab:herx1_sig}
  \begin{tabular}{lcccccccc}
    \toprule
    $E$ (keV)      &
    MDP$_\mathrm{std}$ (\%) &
    MDP$_\mathrm{hyb}$ (\%) &
    R$_\mathrm{MDP}$ &
    $\sigma_\mathrm{std}$ &
    $\sigma_\mathrm{hyb}$ &
    $R$            &
    $(R{-}1)/R_\mathrm{err}$ &
    $\Delta A_\mathrm{eff}/A_\mathrm{eff} (\%)$ \\
    \midrule
    2.00--2.83 & 2.96 & 2.93 & 1.010 & 9.26 &  9.27 & $1.002 \pm 0.002$ & $1.1\sigma$ & $0.4 \pm 0.4$ \\
    2.83--4.00 & 2.28 & 2.24 & 1.018 & 9.29 & 10.18 & $1.096 \pm 0.010$ & $9.3\sigma$ & $20.0 \pm 2.3$ \\
    4.00--5.66 & 2.73 & 2.50 & 1.092 & 9.73 & 10.89 & $1.119 \pm 0.015$ & $7.7\sigma$ & $25.2 \pm 3.4$ \\
    5.66--8.00 & 3.79 & 3.55 & 1.068 & 6.19 &  7.30 & $1.180 \pm 0.029$ & $6.2\sigma$ & $39.3 \pm 6.9$ \\
    \bottomrule
  \end{tabular}
\end{table*}
%
The hybrid MDP is systematically lower than the standard one at all energies because of the higher modulation factor, yielding $R_{\mathrm MDP} > 1$ and $R > 1$ at all energies.
At 2.00-2.83 keV the improvement is consistent with zero at $1.1\sigma$, in agreement with the marginal enhancement predicted at low energies by \citet{2025Cibrario}.
Above 2.83 keV the improvement is highly significant: the significance ratio deviates from unity by $9.3$, $7.7$, and $6.2\sigma$ in the three higher-energy bins, with $R$ reaching $1.18 \pm 0.03$ at 5.66-8.00 keV.
In terms of equivalent effective area (Eq. \ref{eq:DAeff}), the hybrid method provides gains of $20\%{\pm}2\%$, $25\%{\pm}3\%$, and $39\%{\pm}7\%$ in the three bins above 2.83 keV, with a negligible improvement of $0.4\%{\pm}0.4\%$ at 2.00-2.83 keV.
The average improvement in the full 2-8 keV band is of  $9.9\%\pm1.4\%$.

\subsection{Polarimetric results: XTE J1701-461}
\label{subsec:xte}
XTE J1701-461 in its soft state is consistent with zero polarization \citep{2023Cocchi}. 
Fig. \ref{fig:pol_properties} (right column) shows for this source the normalized Stokes parameters, PD, and PA as a function of energy for both methods. 
All values of $q$ and $u$ are consistent with zero within $1$-$2\sigma$ in all energy bins. 
The formally computed PD values are all sub-percent and statistically consistent with zero: they are not significant detections and should formally be reported as upper limits, and PA is unconstrained.
We nonetheless display the computed PD and PA values for both methods to demonstrate their mutual consistency: the two pipelines yield results that agree within uncertainties at all energies. 
In particular, we find no statistically significant differential systematic bias in the polarimetric observables.

\begin{figure}[htbp]
  \centering
  \includegraphics[width=1\linewidth]{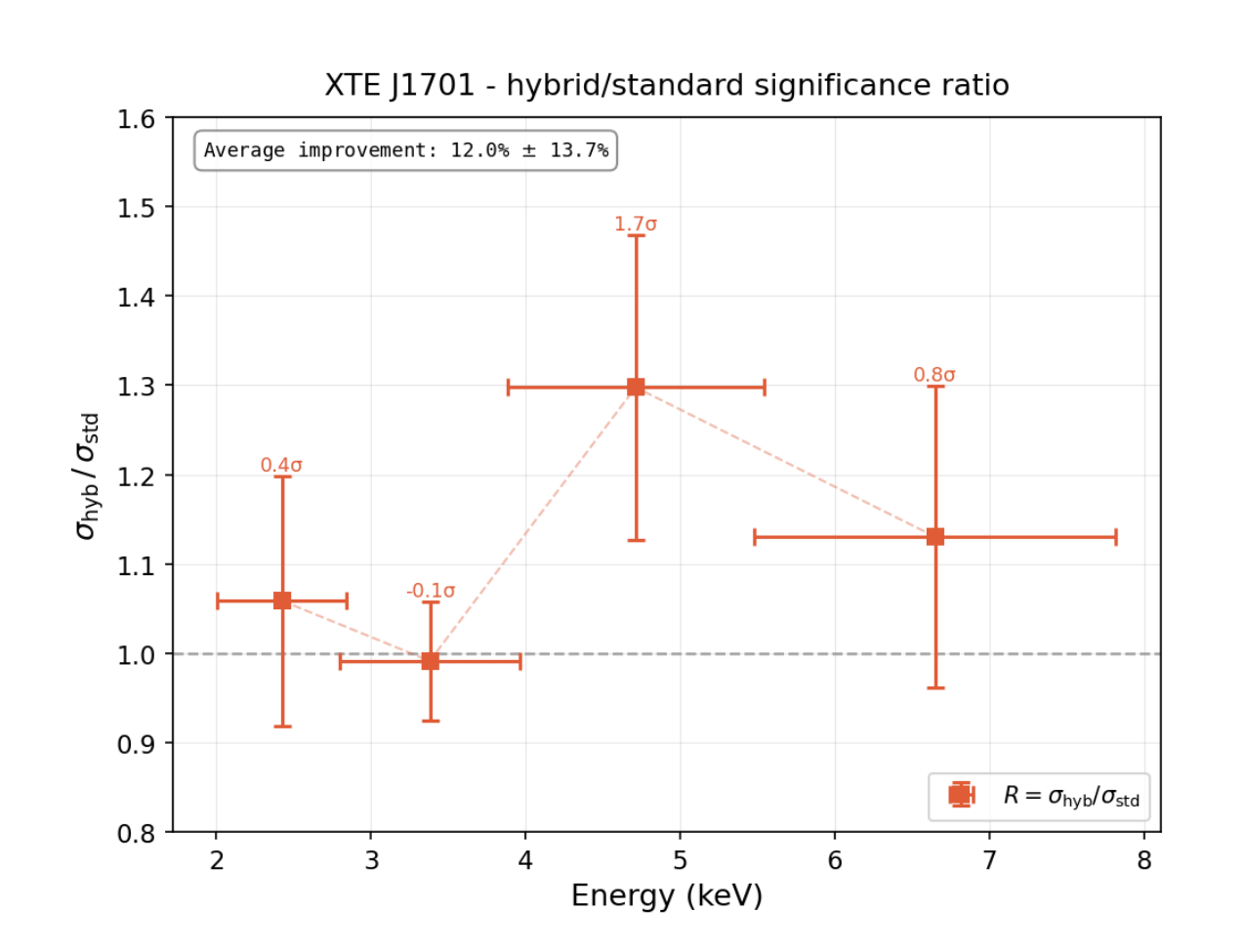}
  \caption{%
    Ratio of detection significances $R = \sigma_\mathrm{hyb}/\sigma_\mathrm{std}$ for XTE J1701-461 as a function of energy. 
    Error bars are computed from Eq.~\eqref{eq:R_err}.
    The dashed horizontal line marks $R = 1$ (equal performance).
    The hybrid method provides no significant improvement over the standard one.}
  \label{fig:significance_ratio_XTE}
\end{figure}

We report the polarimetric parameters in Table \ref{tab:xte_pol} and the significance comparison in Table \ref{tab:xte_sig}.
Fig. \ref{fig:significance_ratio_XTE} shows the significance ratio $R = \sigma_\mathrm{hyb}/\sigma_\mathrm{std}$ as a function of energy.
\begin{table*}[htbp]
  \centering
  \caption{%
    Polarimetric parameters ($q$, $u$, PD, and PA) for XTE J1701-461 as a function of energy, derived with the standard moment analysis (Std) and the hybrid method (Hyb). 
    All quantities are derived with \texttt{PCUBE}.
    None of the PD measurements are statistically significant: PD values are effectively upper limits and PA values are unconstrained (see text).
  }
  \label{tab:xte_pol}
  \setlength{\tabcolsep}{4pt}
  \begin{tabular}{l cc cc cc cc}
    \toprule
    & \multicolumn{2}{c}{$q$ (\%)}
    & \multicolumn{2}{c}{$u$ (\%)}
    & \multicolumn{2}{c}{PD (\%)}
    & \multicolumn{2}{c}{PA (\degr)}\\
    \cmidrule(lr){2-3}\cmidrule(lr){4-5}
    \cmidrule(lr){6-7}\cmidrule(lr){8-9}
    $E$ (keV) & Std & Hyb & Std & Hyb & Std & Hyb & Std & Hyb  \\
    \midrule
    2.00--2.83 & $-0.48{\pm}0.54$ & $-0.60{\pm}0.53$ & $-0.58{\pm}0.54$ & $-0.51{\pm}0.53$ & $0.75{\pm}0.54$ & $0.79{\pm}0.53$ & $-64.6{\pm}20.4$ & $-69.8{\pm}19.2$ \\
    2.83--4.00 & $0.55{\pm}0.40$ & $0.50{\pm}0.39$ & $-0.25{\pm}0.40$ & $-0.29{\pm}0.39$ & $0.60{\pm}0.40$ & $0.58{\pm}0.39$ & $-12.2{\pm}19.1$ & $-15.1{\pm}19.2$ \\
    4.00--5.66 & $-0.38{\pm}0.55$ & $-0.36{\pm}0.50$ & $-0.93{\pm}0.55$ & $-1.14{\pm}0.50$ & $1.00{\pm}0.55$ & $1.19{\pm}0.50$ & $-56.0{\pm}15.7$ & $-53.7{\pm}12.1$ \\
    5.66--8.00 & $-2.43{\pm}1.21$ & $-1.77{\pm}1.12$ & $-1.64{\pm}1.21$ & $-2.52{\pm}1.12$ & $2.93{\pm}1.21$ & $3.08{\pm}1.12$ & $-73.0{\pm}11.8$ & $-62.5{\pm}10.5$ \\
    \bottomrule
  \end{tabular}
\end{table*}
\begin{table*}[htbp]
  \centering
  \caption{%
    Detection significance comparison for XTE J1701-461.
    Energy bins are logarithmically spaced in the 2-8 keV range.
    MDP$_\mathrm{std}$ and MDP$_\mathrm{hyb}$ are the minimum detectable polarizations and R$_\mathrm{MDP}$ their ratio};
    $R = \sigma_\mathrm{hyb}/\sigma_\mathrm{std}$; $R_\mathrm{err}$ is computed
    from Eq. \ref{eq:R_err};
    $(R-1)/R_\mathrm{err}$ is the deviation from parity;
    $\Delta A_\mathrm{eff}/A_\mathrm{eff}$ is the equivalent
    fractional increase in effective area (Eq. \ref{eq:DAeff}).
    All significance values are below $3\sigma$; neither method
    achieves a significant detection.
  \label{tab:xte_sig}
  \begin{tabular}{lcccccccc}
    \toprule
    $E$ (keV)      &
    MDP$_\mathrm{std}$ (\%) &
    MDP$_\mathrm{hyb}$ (\%) &
    R$_\mathrm{MDP}$ &
    $\sigma_\mathrm{std}$ &
    $\sigma_\mathrm{hyb}$ &
    $R$            &
    $(R{-}1)/R_\mathrm{err}$ &
    $\Delta A_\mathrm{eff}/A_\mathrm{eff} (\%)$ \\
    \midrule
    2.00--2.83 & 1.62 & 1.61 & 1.01 &  1.41 &   1.49 &  $1.06 \pm 0.14$ &  $0.4\sigma$ & $12.1 \pm 29.6$ \\
    2.83--4.00 & 1.21 & 1.18 & 1.03 & 1.50 &   1.49  & $0.99 \pm 0.07$ &  $-0.1\sigma$ & $-1.7 \pm 13.2$ \\
    4.00--5.66 & 1.66 & 1.53 & 1.09 & 1.83 &   2.37  & $1.30 \pm 0.17$ &  $1.7\sigma$ & $68.4 \pm 44.2$ \\
    5.66--8.00 & 3.67 & 3.41 & 1.08 & 2.42 &   2.74 &  $1.13 \pm 0.17$ &  $0.8\sigma$ & $27.8 \pm 38.3$ \\
    \bottomrule
  \end{tabular}
\end{table*}
The hybrid method achieves a smaller MDP in all the four energy bins for this source, but provides no significant improvement over the standard one in terms of significance or relative effective area.
None of the individual energy bins resolves $R\neq1$ at more than $1.7\sigma$.
The hybrid method cannot be claimed to provide a significant improvement per-bin for this fainter, lower-significance source ($\sigma_\mathrm{std}\sim1$-$2$, compared to $\sigma_\mathrm{std}\gtrsim9$ for Her X-1).
This is however to be expected for an unpolarized source, so that this result is the astrophysical counterpart of the unpolarized beam analysis of \citet{2025Cibrario} and demonstrates that the hybrid pipeline does not introduce spurious polarization signals for unpolarized sources.
For a source with no significant detection, the most informative figure of merit is the MDP ratio $R_{\rm MDP} = \mathrm{MDP}_{\rm std}/\mathrm{MDP}_{\rm hyb}$ (reported in Table \ref{tab:xte_sig}), which quantifies the sensitivity improvement independently of the realized PD.  
We report $R$ for completeness and to verify that the hybrid pipeline does not artificially inflate the significance for unpolarized sources.
Nonetheless, the central values follow the same trend seen for Her X-1: $R>1$ in three of the four bins, and the 2-8 keV significance increase ($12.0\% \pm13.7\%$) is statistically consistent with the value measured for Her X-1 ($9.9\% \pm 1.4\%$).
This indicates that the larger relative uncertainty for XTE J1701-461 reflects the much lower per-bin significance available in this source ($N_\mathrm{eff}$ and PD both far smaller than for Her X-1), rather than evidence that the hybrid method provides no benefit for fainter or weakly polarized sources; resolving this trend at the bin level for sources in this flux regime would require either longer exposures or a sample of multiple comparably faint sources.

\subsection{Modulation curves and systematics}
\label{subsec:fft_results}
\begin{figure*}[htbp]
  \centering
  \includegraphics[width=1\linewidth]{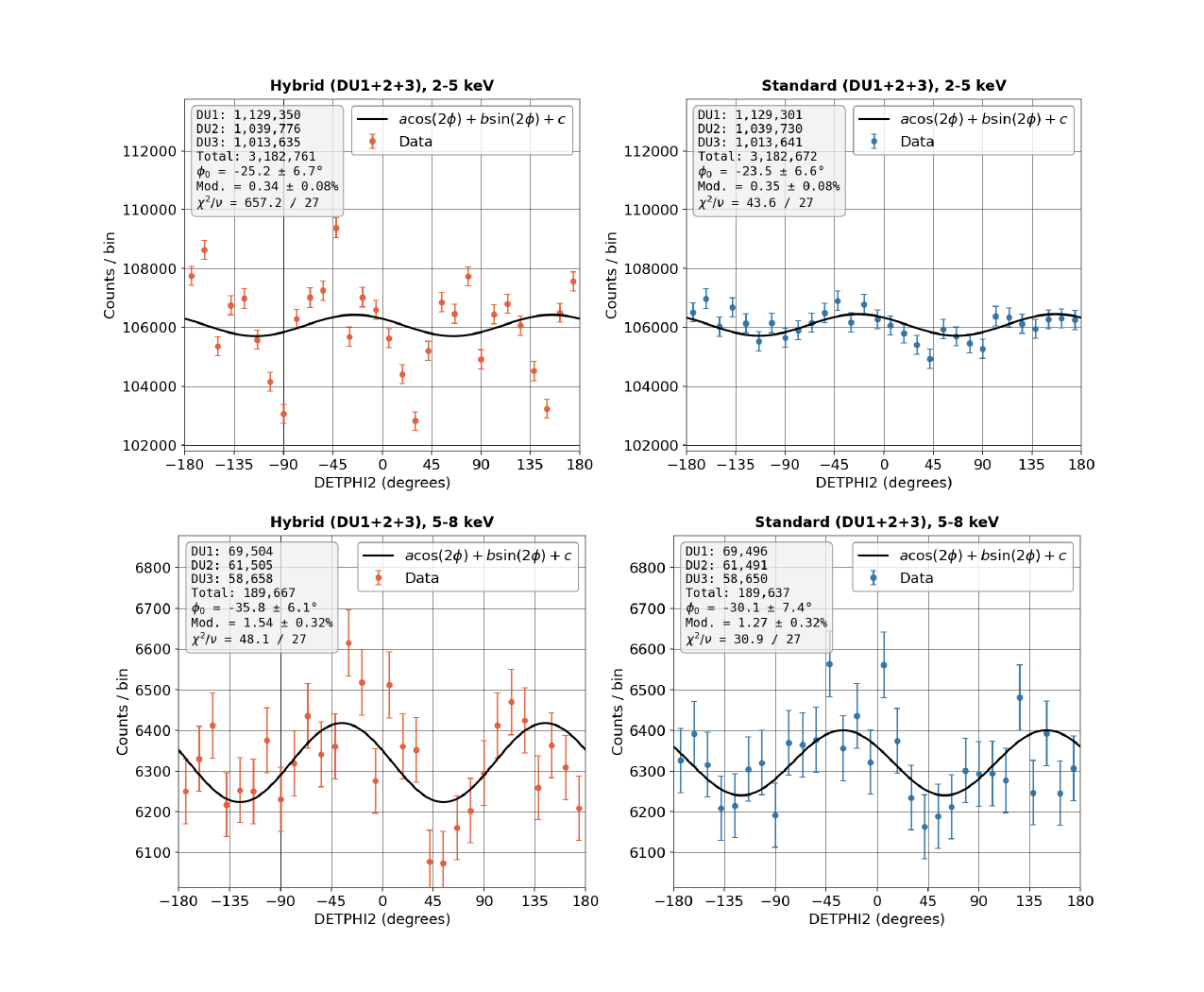}
  \caption{%
    XTE J1701-461 photoelectron emission angle distributions in the 2-5 keV band (top row) and in the 5-8 keV band (bottom row) for the standard (left panels) and hybrid (right panels) methods, summed over the three IXPE detector units after applying the DU rotations. 
    Solid curves show the best-fit modulation model $I(\phi) = a\cos(2\phi) + b\sin(2\phi) + c$; fit parameters and goodness-of-fit are reported in the insets. 
  }
  \label{fig:mod_curve_comparison}
\end{figure*}

\begin{figure*}[htbp!]
  \centering
  \includegraphics[width=0.9\linewidth]{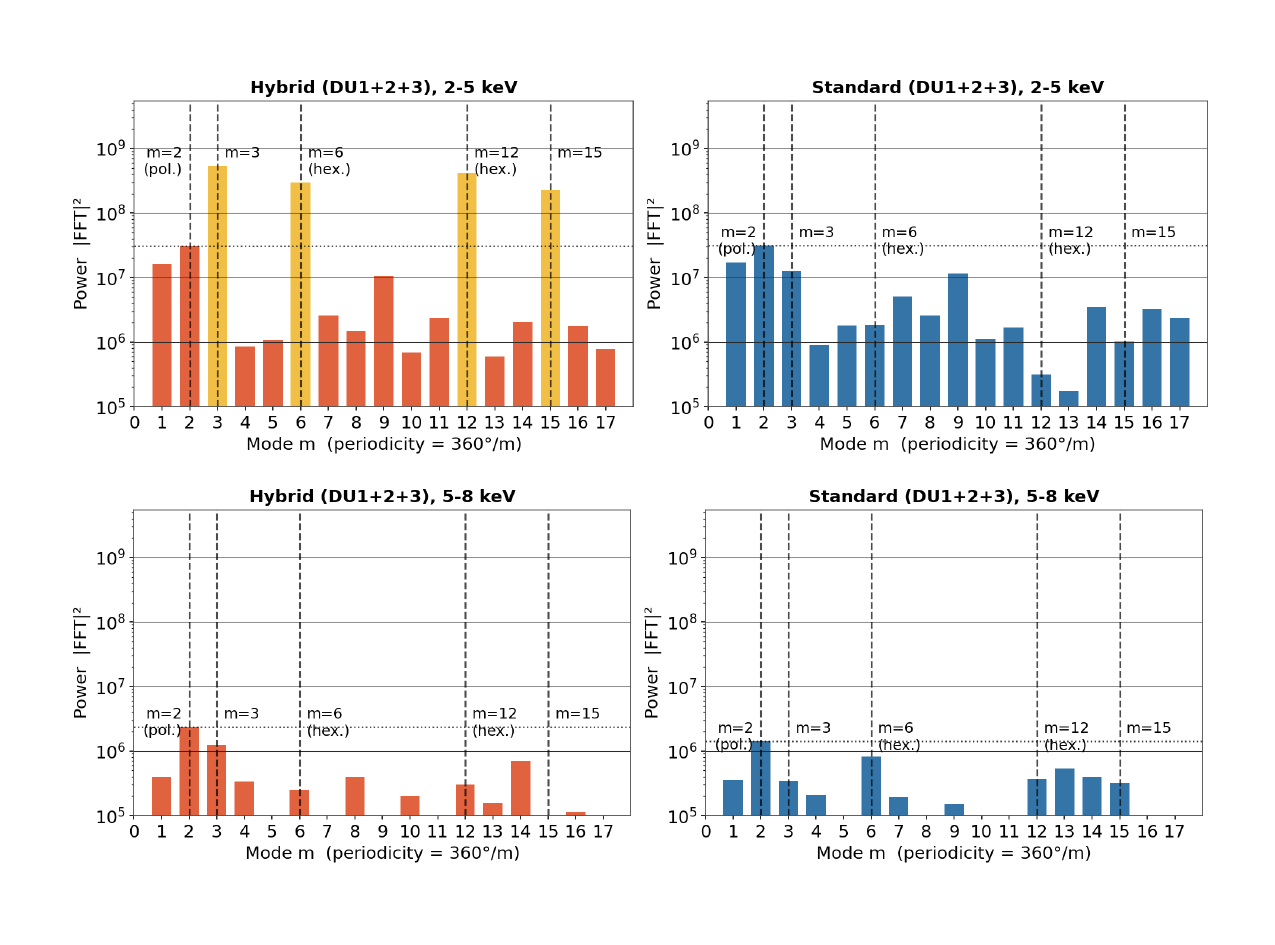}
  \caption{%
    XTE J1701-461 Fourier power spectra of the angle distributions shown in Fig. \ref{fig:mod_curve_comparison} in the 2-5 keV band (top row) and in the 5-8 keV band (bottom row) for the hybrid (left panels) and standard (right panels) methods. 
    The dashed vertical line at $m = 2$ marks the polarization signal, a dotted horizontal line marks its power. 
    Excess power at $m = 3$, $m = 6$, $m = 12$, and $m = 15$ is highlighted in yellow in the hybrid method in the low-energy band but absent in the standard analysis. 
    }
  \label{fig:FFT_comparison}
\end{figure*}

In Fig.\ref{fig:mod_curve_comparison} we show the photoelectron angle distributions and in Fig.\ref{fig:FFT_comparison} their Fourier power spectra in the 2-5 keV and 5-8 keV bands for XTE J1701-461.
   
In the low energy 2-5 keV band we observe excess power at modes at $m = 3$, $m = 6$, $m = 12$, and $m = 15$ in the hybrid angle distributions with respect to the expected polarization mode at $m = 2$.
To quantify the significance of these excesses, we devise a simple statistical test: for a set of $N$ angles drawn from a flat distribution, the Fourier coefficient at each mode $m>0$ is the sum of $N$ unit vectors with random phase, so that its squared modulus (the power) follows an exponential distribution, with mean equal to the underlying noise level of the power spectrum.
We estimate this noise level, separately for each method and energy band, from the sample mean of the power in modes with no expected signal ($m \in \{4,5,7,8,9,10,11,13,14,16\}$), and use it to assign a $p$-value to the power observed at $m = 2, 3, 6, 12$, and $15$, applying a conservative Bonferroni correction for the 15 modes tested \citep[i.e. multiplying the raw $p$-value by 15, ][]{1936Bonferroni}.
Table \ref{tab:fft_significance} reports the resulting significance for both methods and both energy bands.
As expected, the significance of the mode $m=2$, which carries the polarization/spurious-modulation signal, is higher for the hybrid method than for the standard one in both energy bands: this is the same effect, measured here at the level of the individual Fourier mode, that produces the significance ratio $R>1$ for the full PD measurement discussed in Sect.~\ref{subsec:significance} and Sect.~\ref{subsec:herx1}.
\begin{table*}[htbp]
\centering
\caption{Significance of the power excess at modes $m=2,3,6,12,15$ for the standard and hybrid methods, in the 2-5 keV and 5-8 keV bands. $\sigma_\mathrm{hyb}$ and $\sigma_\mathrm{std}$ are the Bonferroni-corrected significances for the 15 modes tested.
"$-$" indicates a corrected significance compatible with pure noise ($p_\mathrm{corr}=1$). $P_\mathrm{hyb}/P_\mathrm{std}$ is the ratio of raw power between the two methods, independent of the noise-level estimate. $|A_m|/N$ is the normalized amplitude of the mode in the hybrid method (2--5~keV), defined as $\sqrt{P_m}/N$ and expressed as a percentage representing the fractional modulation of the angle distribution at that periodicity.}
\label{tab:fft_significance}
\begin{tabular}{c | cccc | ccc}
\toprule
& \multicolumn{4}{c}{2--5 keV} & \multicolumn{3}{c}{5--8 keV} \\
$m$ & $\sigma_\mathrm{hyb}$ & $\sigma_\mathrm{std}$ & $P_\mathrm{hyb}/P_\mathrm{std}$ & $|A_m|/N$ (hyb, \%) & $\sigma_\mathrm{hyb}$ & $\sigma_\mathrm{std}$ & $P_\mathrm{hyb}/P_\mathrm{std}$ \\
\midrule
2  & $5.5$ & $4.3$ & $1.0$    & $0.18$ & $3.4$ & $2.7$  & $1.7$ \\
3  & $7.9$ & $1.7$ & $43.2$   & $0.74$ & $1.7$ & $-$    & $3.7$ \\
6  & $7.9$ & $-$   & $162.2$  & $0.54$ & $-$   & $1.2$  & $0.3$ \\
12 & $7.9$ & $-$   & $1339.8$ & $0.65$ & $-$   & $-$    & $0.8$ \\
15 & $7.9$ & $-$   & $222.0$  & $0.48$ & $-$   & $-$    & $0.1$ \\
\bottomrule
\end{tabular}
\end{table*}
In the 2-5 keV band, the power at $m = 3$, $6$, $12$, and $15$ in the hybrid method exceeds the noise level at significance $>7\sigma$ (Bonferroni-corrected) for all four modes, while the same modes in the standard method are fully consistent with noise (all $<1\sigma$) or only marginally significant.
The ratio of hybrid to standard power in these modes ranges between $\sim$40 and $\sim$1300, confirming that these excesses are a genuine feature of hybrid reconstruction.
However, despite their high statistical significance, the normalized amplitudes $|A_m|/N = \sqrt{P_m}/N$ of all four modes in the hybrid method remain well below 1\% (Table~\ref{tab:fft_significance}), indicating that the modulation of the angle distribution at these periodicities is quantitatively small.
These excesses lead to a markedly poorer fit with the $\cos^2$ modulation model in the hybrid method, as seen from Fig. \ref{fig:mod_curve_comparison}.

On the other hand, in the 5-8 keV band none of $m = 3$, $6$, $12$, or $15$ reaches Bonferroni-corrected significance in either method (Table~\ref{tab:fft_significance}), and the ratio of hybrid to standard power in these modes is of order unity or below. 
This indicates that the systematic excess observed is confined to the energy range where photoelectron tracks are short and only weakly elongated.

As discussed in \citet{2025Cibrario}, the modes $m=6$ and $m=12$ are naturally associated with the hexagonal symmetry of the GPD pixel grid, via the alignment between the barycenter of the horseshoe region selected by the hybrid method and the track barycenter: they are expected to be amplified by the CNN at low energies, where the network tends to predict the IP closer to the track barycenter.
This low-energy regime is the same in which we find the modulation factor improvement of the hybrid method to be negligible (Sec. \ref{subsec:herx1}, $R=1.002\pm0.002$ at 2.00-2.83 keV): since the second moment of the charge distribution about the track barycenter is intrinsically less sensitive to the true emission direction, an IP estimate that remains close to the barycenter carries comparatively little additional polarimetric information relative to the standard method, consistent with both the modulation factor improvement and the $m=6, 12$ excess being most prominent in this same regime.
The excess is reduced in the 5-8 keV energy band, where tracks are longer and the CNN-predicted IP is generally far from the barycenter of the track.

The $m=3$ and $m=15$ modes (corresponding respectively to periodicity of $120\degr$ and $24\degr$, the latter being a harmonic of the former) are not discussed in \citet{2025Cibrario}. 
Both modes are absent in the standard analysis and disappear at 5-8 keV in the hybrid one.
To assess whether these modes could arise from the same horseshoe-barycenter alignment that produces $m = 6$ and $m = 12$, we repeated the moment analysis substituting the CNN-predicted IP with the track barycenter.
This test reproduces significant $m = 6$ and $m = 12$ excesses but yields no excess at $m = 3$ or $m = 15$, ruling out the horseshoe-barycenter alignment as their origin and pointing to a mechanism associated with the CNN inference itself.
A plausible explanation is that they reflect the triple-rotation inference procedure used internally by the hybrid CNN: as described in \citet{2023Cibrario}, for each track the network predicts the IP for three copies of the image rotated by $0\degr$, $120\degr$, and $240\degr$, and averages the three back-rotated predictions.
Small residual asymmetries between these three predictions, more likely for the short and nearly circular tracks found at low energy, could imprint a $120\degr$ periodicity on the angle distribution. 

This procedure has $C_3$ symmetry by construction, and could imprint a $120\degr$ periodicity on the reconstructed angle distribution when the three predictions differ non-negligibly (as expected for the short, nearly circular tracks found at low energies).
Consistently, we verified that $m = 3$ and $m = 15$ excesses are also present in re-analysis of IXPE ground calibration data with the hybrid method, where they had previously gone undetected likely due to the lower signal-to-noise ratio of those measurements relative to the residual spurious modulation and intrinsic beam polarization.
This removes the apparent discrepancy between calibration and astrophysical data, and suggests that the $m = 3$ and $m = 15$ modes are a general feature of the hybrid pipeline rather than a property specific to astrophysical observations.

Crucially, none of these modes have any effect on the polarimetric results.
The normalized Stokes parameters are defined as $q = (2/N)\sum_i \cos(2\phi_i)$ and $u = (2/N)\sum_i \sin(2\phi_i)$ \citep{2015Kislat}: these sums are sensitive only to the $m = 2$ Fourier component of the angle distribution, and all higher-order modes cancel exactly. 
This is demonstrated by the consistency between the two methods in the polarimetric parameters of Sections \ref{subsec:herx1} and \ref{subsec:xte}, which show no evidence of any bias attributable to the modes discussed above.

\section{Discussion and conclusions}
\label{sec:discussion}
 
We presented the first systematic comparison of the hybrid ML-analytic reconstruction method of \citet{2023Cibrario,2025Cibrario} and the standard moment analysis on IXPE data of astrophysical point sources.
 
For the highly polarized source Her X-1, the hybrid method yields a consistently higher detection significance at all energies above $\sim$2.83~keV, with improvements that are statistically significant at the level of 5.8--9.3$\sigma$ in three of four energy bins. 
The improvement is negligible at 2.00-2.83 keV ($1.1\sigma$), consistent with the predictions of \citet{2025Cibrario}: at low energies the photoelectron tracks are nearly circular, providing little morphological information to improve the IP reconstruction. 
Above $\sim$3 keV, where the tracks become elongated and asymmetric, the IP predicted by the CNN is significantly more accurate, leading to improved polarimetric sensitivity.
 
The magnitude of the improvement is up to $\sim$18\% in $\sigma_\mathrm{hyb}/\sigma_\mathrm{std}$ at 5.66-8.00~keV. 
For bright sources near the minimum detectable polarization limit, an improvement of this magnitude can make the difference between a detection and an upper limit.
In terms of the equivalent fractional increase in the effective area $\Delta A_\mathrm{eff}/A_\mathrm{eff} = R^2 - 1$
(Eq. \eqref{eq:DAeff}), it ranges from $20\%$ at 2.83-4.00 keV to $39\%$ at 5.66-8.00 keV at no cost of additional observation time.
 
For the observationally unpolarized source XTE J1701-461, the per-bin significance ratios are all consistent with unity within their uncertainties, as expected given the much lower per-bin significance available for this source ($\sigma_\mathrm{std}\sim1$--$2$) compared to Her X-1 ($\sigma_\mathrm{std}\gtrsim9$).
The source-averaged central value ($12.0\%\pm13.7\%$) is nonetheless consistent with the trend established on Her X-1 ($9.9\%\pm1.4\%$), and the hybrid method achieves a smaller PD uncertainty than the standard one in three of the four energy bins (Table~\ref{tab:xte_pol}). 
This indicates that the present non-detection of a per-bin improvement for XTE J1701-461 reflects the limited statistics of this fainter source rather than an intrinsic loss of benefit from the hybrid method at low source flux or polarization.
We stress that this conclusion is established here for bright point sources only, and tests similar to the ones performed in this work should be performed for any applications of this NN to other data sets.

Both sources analyzed in this work have per-bin significance $\sigma_\mathrm{std}\gtrsim1$, and the benefit of the hybrid method is robustly detected only for the brighter of the two; for the unpolarized XTE J1701-461, the same central improvement is present, but not individually resolved in the available statistics (Sect.~\ref{subsec:xte}).

The central PD values show a mild tendency to be larger for the hybrid reconstruction: across the eight energy bins of the two sources, $\mathrm{PD}{\rm hyb}>\mathrm{PD}{\rm std}$ in six cases. 
However, this does not constitute statistically significant evidence for a systematic positive offset. 
A binomial sign test gives a one-sided probability of 14.5\% for obtaining six or more positive differences under the null hypothesis of no preferred direction. 
Excluding the lowest-energy bin of each source leaves five positive differences out of six, with a corresponding one-sided probability of 10.9\%. 
We therefore interpret this pattern as a possible realization-dependent fluctuation rather than evidence for a differential systematic bias.
The observed significance ratio contains this effect explicitly. 
For example, using the values in the 2.83-4.00 keV bin of HerX-1 from Table \ref{tab:herx1_pol},
\begin{equation}
R_{\rm PD}=\frac{7.66}{7.13}=1.074,
\qquad
R_{\rm err}=\frac{0.77}{0.75}=1.027,
\end{equation}
so that the observed value $R=1.096$ contains contributions from both the reduced PD uncertainty and the higher central PD measured with the hybrid reconstruction. 
If the two methods had returned the same PD, the significance ratio would instead have been $R=1.027$.
The MDP ratio R$_{\mathrm{MDP}}$ provides a complementary measure that is independent of this particular PD realization. 
For Her X-1, R$_{\mathrm{MDP}}$ is 1.010, 1.018, 1.092, and 1.068 in the four energy bins, confirming a sensitivity improvement consistent with that expected from the modulation-factor enhancement of the hybrid reconstruction.\\
Preliminary comparisons on faint extended sources, such as supernova remnants, show no measurable improvement with the hybrid method, with results from the two pipelines being statistically consistent but offering no significant gain in sensitivity.
This suggests the existence of a flux (or, more precisely, a per-bin significance) threshold below which the benefit of the improved IP reconstruction is not recoverable, plausibly because the gain in $N_\mathrm{eff}$ and $\mu$ documented here for Her X-1 becomes comparable to or smaller than the photon-counting noise once the number of source counts is sufficiently low.
A quantitative determination of this threshold, and an explanation of why the method does not yet show a benefit for fainter sources, requires a dedicated analysis of a larger source sample spanning a range of fluxes and is left to future work.
All in all, the comparison of the analysis of both sources shows no statistically significant evidence for a differential systematic bias between the two reconstruction methods at the present statistical precision. We nevertheless note a mild tendency for the hybrid reconstruction to yield larger central PD values in the present sample. Given the limited number of independent measurements, this tendency cannot be distinguished from a statistical fluctuation, although a small residual method-dependent contribution cannot be excluded.
The Fourier analysis of the level 1 angle distributions reveals the power excess $m = 6$, $m = 12$, $m = 3$ and $m = 15$ in the hybrid method at $2-5$ keV, significant at $7.9\sigma$ for all four modes, compared to a standard method that is either consistent with noise or only marginally significant for the same modes (Table \ref{tab:fft_significance}); at $5-8$ keV none of these modes reach significance in either method. 
This confirms that the excess is a genuine, energy-dependent systematic of the hybrid reconstruction, and one that is fully characterized by the comparative test presented here, regardless of its ultimate physical origin. 
A barycenter-as-IP test confirms that $m = 6$ and $m = 12$ originate from the horseshoe-barycenter alignment described in \citet{2025Cibrario}, while $m = 3$ and $m = 15$ require a distinct mechanism, plausibly the $C_3$-symmetric triple-rotation inference procedure of the CNN. 
We further verified that $m = 3$ and $m = 15$ are present in re-analysis of IXPE ground calibration data with the hybrid method, confirming that they are a general feature of the hybrid pipeline.
We confirm that none of these modes propagates to the Stokes parameters, which is consistent with the analysis of \citet{2025Cibrario}.
We note that a similar $120\degr$ periodicity artifact has been independently reported for a different CNN-based reconstruction method applied to IXPE ground calibration data: \citet{2022a_DiMarco} find that the deep-ensemble approach of \citet{2021Peirson} introduces $m=6$ and $m=12$ harmonics in the modulation curves of polarized calibration data, which they attribute to the network's processing of the hexagonal track image, itself based on three $120\degr$ separated rotations to map the hexagonal grid onto a square one \citep[Sect. 2.2]{2021Peirson}. 
This suggests that $120\degr$ periodicity artifacts may be a generic consequence of CNN-based approaches that explicitly exploit the threefold rotational structure of the GPD hexagonal grid, rather than a peculiarity of the specific architecture used by \citet{2023Cibrario}.
Recently, a method using a Graph Neural Network (GNN) trained on null polarization signals successfully resolved the harmonics implied by the hexagonal pattern of the IXPE detector, but without a significant improvement over the standard method in terms of polarimetric performance \citep{latorre2026}. 

In summary, the hybrid reconstruction method is ready for a broader application to IXPE astrophysical data of bright point sources. 
Its adoption as the standard analysis pipeline in the $\gtrsim$3~keV band would provide a sensitivity improvement equivalent to an average increase in the effective area of $\sim20\%$ in the entire 2 - 8 keV IXPE energy band, without additional observational cost.

\begin{acknowledgements}
The Imaging X-ray Polarimetry Explorer (IXPE) is a joint US and Italian mission.  
The US contribution is supported by the National Aeronautics and Space Administration (NASA) and led and managed by its Marshall Space Flight Center (MSFC), with industry partner Ball Aerospace (contract NNM15AA18C). 
The Italian contribution is supported by the Italian Space Agency (Agenzia Spaziale Italiana, ASI) through contract ASI-OHBI-2022-13-I.0, agreements ASI-INAF-2022-19-HH.0 and ASI-INFN-2017.13-H0, and its Space Science Data Center (SSDC) with agreements ASI-INAF-2022-14-HH.0 and ASI-INFN 2021-43-HH.0, and by the Istituto Nazionale di Astrofisica (INAF) and the Istituto Nazionale di Fisica Nucleare (INFN) in Italy.  
This research used data products provided by the IXPE Team (MSFC, SSDC, INAF, and INFN) and distributed with additional software tools by the High-Energy Astrophysics Science Archive Research Center (HEASARC), at NASA Goddard Space Flight Center (GSFC).
NC and MN wish to thank Nikita Moriakov for his contribution to the original implementation of hybrid method, first presented in \cite{2023Cibrario}. 
\end{acknowledgements}

%

\bibliography{biblio} 

@ARTICLE{2006Vaillancourt,
       author = {{Vaillancourt}, John E.},
        title = "{Placing Confidence Limits on Polarization Measurements}",
      journal = {\pasp},
         year = 2006,
        month = sep,
       volume = {118},
       number = {847},
        pages = {1340-1343},
          doi = {10.1086/507472},
archivePrefix = {arXiv},
       eprint = {astro-ph/0603110},
 primaryClass = {astro-ph},
       adsurl = {https://ui.adsabs.harvard.edu/abs/2006PASP..118.1340V}
}

@ARTICLE{2015Kislat,
       author = {{Kislat}, F. and {Clark}, B. and {Beilicke}, M. and {Krawczynski}, H.},
        title = "{Analyzing the data from X-ray polarimeters with Stokes parameters}",
      journal = {Astroparticle Physics},
         year = 2015,
        month = aug,
       volume = {68},
        pages = {45-51},
          doi = {10.1016/j.astropartphys.2015.02.007},
archivePrefix = {arXiv},
       eprint = {1409.6214},
 primaryClass = {astro-ph.IM},
       adsurl = {https://ui.adsabs.harvard.edu/abs/2015APh....68...45K}
}

@ARTICLE{2022Weisskopf,
       author = {{Weisskopf}, Martin C. and {Soffitta}, Paolo and {Baldini}, Luca and {Ramsey}, Brian D. and {O'Dell}, Stephen L. and {Romani}, Roger W. and {Matt}, Giorgio and {Deininger}, William D. and {Baumgartner}, Wayne H. and {Bellazzini}, Ronaldo and {Costa}, Enrico and {Kolodziejczak}, Jeffery J. and {Latronico}, Luca and {Marshall}, Herman L. and {Muleri}, Fabio and {Bongiorno}, Stephen D. and {Tennant}, Allyn and {Bucciantini}, Niccolo and {Dovciak}, Michal and {Marin}, Frederic and {Marscher}, Alan and {Poutanen}, Juri and {Slane}, Pat and {Turolla}, Roberto and {Kalinowski}, William and {Di Marco}, Alessandro and {Fabiani}, Sergio and {Minuti}, Massimo and {La Monaca}, Fabio and {Pinchera}, Michele and {Rankin}, John and {Sgro'}, Carmelo and {Trois}, Alessio and {Xie}, Fei and {Alexander}, Cheryl and {Allen}, D. Zachery and {Amici}, Fabrizio and {Andersen}, Jason and {Antonelli}, Angelo and {Antoniak}, Spencer and {Attin{\`a}}, Primo and {Barbanera}, Mattia and {Bachetti}, Matteo and {Baggett}, Randy M. and {Bladt}, Jeff and {Brez}, Alessandro and {Bonino}, Raffaella and {Boree}, Christopher and {Borotto}, Fabio and {Breeding}, Shawn and {Brienza}, Daniele and {Bygott}, H. Kyle and {Caporale}, Ciro and {Cardelli}, Claudia and {Carpentiero}, Rita and {Castellano}, Simone and {Castronuovo}, Marco and {Cavalli}, Luca and {Cavazzuti}, Elisabetta and {Ceccanti}, Marco and {Centrone}, Mauro and {Citraro}, Saverio and {D'Amico}, Fabio and {D'Alba}, Elisa and {Di Gesu}, Laura and {Del Monte}, Ettore and {Dietz}, Kurtis L. and {Di Lalla}, Niccolo' and {Persio}, Giuseppe Di and {Dolan}, David and {Donnarumma}, Immacolata and {Evangelista}, Yuri and {Ferrant}, Kevin and {Ferrazzoli}, Riccardo and {Ferrie}, MacKenzie and {Footdale}, Joseph and {Forsyth}, Brent and {Foster}, Michelle and {Garelick}, Benjamin and {Gunji}, Shuichi and {Gurnee}, Eli and {Head}, Michael and {Hibbard}, Grant and {Johnson}, Samantha and {Kelly}, Erik and {Kilaru}, Kiranmayee and {Lefevre}, Carlo and {Roy}, Shelley Le and {Loffredo}, Pasqualino and {Lorenzi}, Paolo and {Lucchesi}, Leonardo and {Maddox}, Tyler and {Magazzu}, Guido and {Maldera}, Simone and {Manfreda}, Alberto and {Mangraviti}, Elio and {Marengo}, Marco and {Marrocchesi}, Alessandra and {Massaro}, Francesco and {Mauger}, David and {McCracken}, Jeffrey and {McEachen}, Michael and {Mize}, Rondal and {Mereu}, Paolo and {Mitchell}, Scott and {Mitsuishi}, Ikuyuki and {Morbidini}, Alfredo and {Mosti}, Federico and {Nasimi}, Hikmat and {Negri}, Barbara and {Negro}, Michela and {Nguyen}, Toan and {Nitschke}, Isaac and {Nuti}, Alessio and {Onizuka}, Mitch and {Oppedisano}, Chiara and {Orsini}, Leonardo and {Osborne}, Darren and {Pacheco}, Richard and {Paggi}, Alessandro and {Painter}, Will and {Pavelitz}, Steven D. and {Pentz}, Christina and {Piazzolla}, Raffaele and {Perri}, Matteo and {Pesce-Rollins}, Melissa and {Peterson}, Colin and {Pilia}, Maura and {Profeti}, Alessandro and {Puccetti}, Simonetta and {Ranganathan}, Jaganathan and {Ratheesh}, Ajay and {Reedy}, Lee and {Root}, Noah and {Rubini}, Alda and {Ruswick}, Stephanie and {Sanchez}, Javier and {Sarra}, Paolo and {Santoli}, Francesco and {Scalise}, Emanuele and {Sciortino}, Andrea and {Schroeder}, Christopher and {Seek}, Tim and {Sosdian}, Kalie and {Spandre}, Gloria and {Speegle}, Chet O. and {Tamagawa}, Toru and {Tardiola}, Marcello and {Tobia}, Antonino and {Thomas}, Nicholas E. and {Valerie}, Robert and {Vimercati}, Marco and {Walden}, Amy L. and {Weddendorf}, Bruce and {Wedmore}, Jeffrey and {Welch}, David and {Zanetti}, Davide and {Zanetti}, Francesco},
        title = "{The Imaging X-Ray Polarimetry Explorer (IXPE): Pre-Launch}",
      journal = {Journal of Astronomical Telescopes, Instruments, and Systems},
         year = 2022,
        month = apr,
       volume = {8},
       number = {2},
          eid = {026002},
        pages = {026002},
          doi = {10.1117/1.JATIS.8.2.026002},
archivePrefix = {arXiv},
       eprint = {2112.01269},
 primaryClass = {astro-ph.IM},
       adsurl = {https://ui.adsabs.harvard.edu/abs/2022JATIS...8b6002W}
}

@ARTICLE{2021Soffitta,
       author = {{Soffitta}, Paolo and {Baldini}, Luca and {Bellazzini}, Ronaldo and {Costa}, Enrico and {Latronico}, Luca and {Muleri}, Fabio and {Del Monte}, Ettore and {Fabiani}, Sergio and {Minuti}, Massimo and {Pinchera}, Michele and {Sgro'}, Carmelo and {Spandre}, Gloria and {Trois}, Alessio and {Amici}, Fabrizio and {Andersson}, Hans and {Attina'}, Primo and {Bachetti}, Matteo and {Barbanera}, Mattia and {Borotto}, Fabio and {Brez}, Alessandro and {Brienza}, Daniele and {Caporale}, Ciro and {Cardelli}, Claudia and {Carpentiero}, Rita and {Castellano}, Simone and {Castronuovo}, Marco and {Cavalli}, Luca and {Cavazzuti}, Elisabetta and {Ceccanti}, Marco and {Centrone}, Mauro and {Ciprini}, Stefano and {Citraro}, Saverio and {D'Amico}, Fabio and {D'Alba}, Elisa and {Di Cosimo}, Sergio and {Di Lalla}, Niccolo' and {Di Marco}, Alessandro and {Di Persio}, Giuseppe and {Donnarumma}, Immacolata and {Evangelista}, Yuri and {Ferrazzoli}, Riccardo and {Hayato}, Asami and {Kitaguchi}, Takao and {La Monaca}, Fabio and {Lefevre}, Carlo and {Loffredo}, Pasqualino and {Lorenzi}, Paolo and {Lucchesi}, Leonardo and {Magazzu}, Carlo and {Maldera}, Simone and {Manfreda}, Alberto and {Mangraviti}, Elio and {Marengo}, Marco and {Matt}, Giorgio and {Mereu}, Paolo and {Morbidini}, Alfredo and {Mosti}, Federico and {Nakano}, Toshio and {Nasimi}, Hikmat and {Negri}, Barbara and {Nenonen}, Seppo and {Nuti}, Alessio and {Orsini}, Leonardo and {Perri}, Matteo and {Pesce-Rollins}, Melissa and {Piazzolla}, Raffaele and {Pilia}, Maura and {Profeti}, Alessandro and {Puccetti}, Simonetta and {Rankin}, John and {Ratheesh}, Ajay and {Rubini}, Alda and {Santoli}, Francesco and {Sarra}, Paolo and {Scalise}, Emanuele and {Sciortino}, Andrea and {Tamagawa}, Toru and {Tardiola}, Marcello and {Tobia}, Antonino and {Vimercati}, Marco and {Xie}, Fei},
        title = "{The Instrument of the Imaging X-Ray Polarimetry Explorer}",
      journal = {\aj},
         year = 2021,
        month = nov,
       volume = {162},
       number = {5},
          eid = {208},
        pages = {208},
          doi = {10.3847/1538-3881/ac19b0},
archivePrefix = {arXiv},
       eprint = {2108.00284},
 primaryClass = {astro-ph.IM},
       adsurl = {https://ui.adsabs.harvard.edu/abs/2021AJ....162..208S}
}

@ARTICLE{2021Baldini,
       author = {{Baldini}, L. and {Barbanera}, M. and {Bellazzini}, R. and {Bonino}, R. and {Borotto}, F. and {Brez}, A. and {Caporale}, C. and {Cardelli}, C. and {Castellano}, S. and {Ceccanti}, M. and {Citraro}, S. and {Di Lalla}, N. and {Latronico}, L. and {Lucchesi}, L. and {Magazz{\`u}}, C. and {Magazz{\`u}}, G. and {Maldera}, S. and {Manfreda}, A. and {Marengo}, M. and {Marrocchesi}, A. and {Mereu}, P. and {Minuti}, M. and {Mosti}, F. and {Nasimi}, H. and {Nuti}, A. and {Oppedisano}, C. and {Orsini}, L. and {Pesce-Rollins}, M. and {Pinchera}, M. and {Profeti}, A. and {Sgr{\`o}}, C. and {Spandre}, G. and {Tardiola}, M. and {Zanetti}, D. and {Amici}, F. and {Andersson}, H. and {Attin{\`a}}, P. and {Bachetti}, M. and {Baumgartner}, W. and {Brienza}, D. and {Carpentiero}, R. and {Castronuovo}, M. and {Cavalli}, L. and {Cavazzuti}, E. and {Centrone}, M. and {Costa}, E. and {D'Alba}, E. and {D'Amico}, F. and {Del Monte}, E. and {Di Cosimo}, S. and {Di Marco}, A. and {Di Persio}, G. and {Donnarumma}, I. and {Evangelista}, Y. and {Fabiani}, S. and {Ferrazzoli}, R. and {Kitaguchi}, T. and {La Monaca}, F. and {Lefevre}, C. and {Loffredo}, P. and {Lorenzi}, P. and {Mangraviti}, E. and {Matt}, G. and {Meilahti}, T. and {Morbidini}, A. and {Muleri}, F. and {Nakano}, T. and {Negri}, B. and {Nenonen}, S. and {O'Dell}, S.~L. and {Perri}, M. and {Piazzolla}, R. and {Pieraccini}, S. and {Pilia}, M. and {Puccetti}, S. and {Ramsey}, B.~D. and {Rankin}, J. and {Ratheesh}, A. and {Rubini}, A. and {Santoli}, F. and {Sarra}, P. and {Scalise}, E. and {Sciortino}, A. and {Soffitta}, P. and {Tamagawa}, T. and {Tennant}, A.~F. and {Tobia}, A. and {Trois}, A. and {Uchiyama}, K. and {Vimercati}, M. and {Weisskopf}, M.~C. and {Xie}, F. and {Zanetti}, F. and {Zhou}, Y.},
        title = "{Design, construction, and test of the Gas Pixel Detectors for the IXPE mission}",
      journal = {Astroparticle Physics},
         year = 2021,
        month = dec,
       volume = {133},
          eid = {102628},
        pages = {102628},
          doi = {10.1016/j.astropartphys.2021.102628},
archivePrefix = {arXiv},
       eprint = {2107.05496},
 primaryClass = {astro-ph.IM},
       adsurl = {https://ui.adsabs.harvard.edu/abs/2021APh...13302628B}
}

@ARTICLE{2001Costa,
       author = {{Costa}, Enrico and {Soffitta}, Paolo and {Bellazzini}, Ronaldo and {Brez}, Alessandro and {Lumb}, Nicholas and {Spandre}, Gloria},
        title = "{An efficient photoelectric X-ray polarimeter for the study of black holes and neutron stars}",
      journal = {\nat},
         year = 2001,
        month = jun,
       volume = {411},
       number = {6838},
        pages = {662-665},
          doi = {10.1038/35079508},
archivePrefix = {arXiv},
       eprint = {astro-ph/0107486},
 primaryClass = {astro-ph},
       adsurl = {https://ui.adsabs.harvard.edu/abs/2001Natur.411..662C}
}

@ARTICLE{2006Bellazzini,
       author = {{Bellazzini}, R. and {Angelini}, F. and {Baldini}, L. and {Bitti}, F. and {Brez}, A. and {Cavalca}, F. and {Del Prete}, M. and {Kuss}, M. and {Latronico}, L. and {Omodei}, N. and {Pinchera}, M. and {Massai}, M.~M. and {Minuti}, M. and {Razzano}, M. and {Sgro}, C. and {Spandre}, G. and {Tenze}, A. and {Costa}, E. and {Soffitta}, P.},
        title = "{Gas pixel detectors for X-ray polarimetry applications}",
      journal = {Nuclear Instruments and Methods in Physics Research A},
         year = 2006,
        month = may,
       volume = {560},
       number = {2},
        pages = {425-434},
          doi = {10.1016/j.nima.2006.01.046},
archivePrefix = {arXiv},
       eprint = {astro-ph/0512242},
 primaryClass = {astro-ph},
       adsurl = {https://ui.adsabs.harvard.edu/abs/2006NIMPA.560..425B}
}

@ARTICLE{2023Cibrario,
       author = {{Cibrario}, N. and {Negro}, M. and {Moriakov}, N. and {Bonino}, R. and {Baldini}, L. and {Di Lalla}, N. and {Latronico}, L. and {Maldera}, S. and {Manfreda}, A. and {Omodei}, N. and {Sgr{\'o}}, C. and {Tugliani}, S.},
        title = "{Joint machine learning and analytic track reconstruction for X-ray polarimetry with gas pixel detectors}",
      journal = {\aap},
         year = 2023,
        month = jun,
       volume = {674},
          eid = {A107},
        pages = {A107},
          doi = {10.1051/0004-6361/202346302},
archivePrefix = {arXiv},
       eprint = {2304.14425},
 primaryClass = {astro-ph.IM},
       adsurl = {https://ui.adsabs.harvard.edu/abs/2023A&A...674A.107C}
}

@ARTICLE{2025Cibrario,
       author = {{Cibrario}, Nicol{\'o} and {Negro}, Michela and {Bonino}, Raffaella and {Moriakov}, Nikita and {Baldini}, Luca and {Di Lalla}, Niccol{\'o} and {Di Marco}, Alessandro and {Fabiani}, Sergio and {Frass{\'a}}, Andrea and {Gorgi}, Alessio and {La Monaca}, Fabio and {Latronico}, Luca and {Maldera}, Simone and {Manfreda}, Alberto and {Muleri}, Fabio and {Omodei}, Nicola and {Rankin}, John and {Sgr{\'o}}, Carmelo and {Silvestri}, Stefano and {Soffitta}, Paolo and {Tugliani}, Stefano},
        title = "{Mitigating Polarization Leakage in Gas Pixel Detectors through Hybrid Machine Learning and Analytic Event Reconstruction}",
      journal = {\apj},
         year = 2025,
        month = may,
       volume = {984},
       number = {2},
          eid = {171},
        pages = {171},
          doi = {10.3847/1538-4357/adc92c},
archivePrefix = {arXiv},
       eprint = {2506.07828},
 primaryClass = {astro-ph.IM},
       adsurl = {https://ui.adsabs.harvard.edu/abs/2025ApJ...984..171C}
}

@ARTICLE{2023Bucciantini_a,
       author = {{Bucciantini}, N. and {Di Lalla}, N. and {Romani}, R.~W.~R. and {Silvestri}, S. and {Negro}, M. and {Baldini}, L. and {Tennant}, A.~F. and {Manfreda}, A.},
        title = "{Polarisation leakage due to errors in track reconstruction in gas pixel detectors}",
      journal = {\aap},
         year = 2023,
        month = apr,
       volume = {672},
          eid = {A66},
        pages = {A66},
          doi = {10.1051/0004-6361/202245744},
archivePrefix = {arXiv},
       eprint = {2302.00346},
 primaryClass = {astro-ph.IM},
       adsurl = {https://ui.adsabs.harvard.edu/abs/2023A&A...672A..66B}
}

@INPROCEEDINGS{2003Bellazzini,
       author = {{Bellazzini}, Ronaldo and {Angelini}, F. and {Baldini}, Luca and {Brez}, Alessandro and {Costa}, Enrico and {Di Persio}, Giuseppe and {Latronico}, Luca and {Massai}, M.~M. and {Omodei}, Nicola and {Pacciani}, Luigi and {Soffitta}, Paolo and {Spandre}, Gloria},
        title = "{Novel gaseus X-ray polarimeter: data analysis and simulation}",
    booktitle = {Polarimetry in Astronomy},
         year = 2003,
       editor = {{Fineschi}, Silvano},
       series = {Society of Photo-Optical Instrumentation Engineers (SPIE) Conference Series},
       volume = {4843},
        month = feb,
        pages = {383-393},
          doi = {10.1117/12.459381},
       adsurl = {https://ui.adsabs.harvard.edu/abs/2003SPIE.4843..383B}
}

@ARTICLE{2016Huang,
       author = {{Huang}, Gao and {Liu}, Zhuang and {van der Maaten}, Laurens and {Weinberger}, Kilian Q.},
        title = "{Densely Connected Convolutional Networks}",
      journal = {arXiv e-prints},
         year = 2016,
        month = aug,
          eid = {arXiv:1608.06993},
        pages = {arXiv:1608.06993},
          doi = {10.48550/arXiv.1608.06993},
archivePrefix = {arXiv},
       eprint = {1608.06993},
 primaryClass = {cs.CV},
       adsurl = {https://ui.adsabs.harvard.edu/abs/2016arXiv160806993H}
}

@ARTICLE{2019Steppa,
       author = {{Steppa}, Constantin and {Holch}, Tim L.},
        title = "{HexagDLy-Processing hexagonally sampled data with CNNs in PyTorch}",
      journal = {SoftwareX},
         year = 2019,
        month = jan,
       volume = {9},
        pages = {193-198},
          doi = {10.1016/j.softx.2019.02.010},
archivePrefix = {arXiv},
       eprint = {1903.01814},
 primaryClass = {cs.CV},
       adsurl = {https://ui.adsabs.harvard.edu/abs/2019SoftX...9..193S}
}

@ARTICLE{2022Muleri,
       author = {{Muleri}, Fabio and {Piazzolla}, Raffaele and {Di Marco}, Alessandro and {Fabiani}, Sergio and {La Monaca}, Fabio and {Lefevre}, Carlo and {Morbidini}, Alfredo and {Rankin}, John and {Soffitta}, Paolo and {Tobia}, Antonino and {Xie}, Fei and {Amici}, Fabrizio and {Attin{\`a}}, Primo and {Bachetti}, Matteo and {Brienza}, Daniele and {Centrone}, Mauro and {Costa}, Enrico and {Del Monte}, Ettore and {Di Cosimo}, Sergio and {Di Persio}, Giuseppe and {Evangelista}, Yuri and {Ferrazzoli}, Riccardo and {Loffredo}, Pasqualino and {Perri}, Matteo and {Pilia}, Maura and {Puccetti}, Simonetta and {Ratheesh}, Ajay and {Rubini}, Alda and {Santoli}, Francesco and {Scalise}, Emanuele and {Trois}, Alessio},
        title = "{The IXPE instrument calibration equipment}",
      journal = {Astroparticle Physics},
         year = 2022,
        month = mar,
       volume = {136},
          eid = {102658},
        pages = {102658},
          doi = {10.1016/j.astropartphys.2021.102658},
archivePrefix = {arXiv},
       eprint = {2111.02066},
 primaryClass = {astro-ph.IM},
       adsurl = {https://ui.adsabs.harvard.edu/abs/2022APh...13602658M}
}

@INPROCEEDINGS{2025DiLalla,
       author = {{Di Lalla}, Niccolo' and {Omodei}, Nicola and {Bucciantini}, Niccolo},
        title = "{X-ray Polarization Detection of the Pulsar Wind Nebula in G21.5-0.9 with IXPE}",
    booktitle = {246th Meeting of the American Astronomical Society},
         year = 2025,
       series = {American Astronomical Society Meeting Abstracts},
       volume = {246},
        month = jun,
          eid = {117.06},
        pages = {117.06},
       adsurl = {https://ui.adsabs.harvard.edu/abs/2025AAS...24611706D}
}

@ARTICLE{2022DiMarco,
       author = {{Di Marco}, Alessandro and {Fabiani}, Sergio and {La Monaca}, Fabio and {Muleri}, Fabio and {Rankin}, John and {Soffitta}, Paolo and {Xie}, Fei and {Amici}, Fabrizio and {attin{\`a}}, Primo and {Bachetti}, Matteo and {Baldini}, Luca and {Barbanera}, Mattia and {Baumgartner}, Wayne and {Bellazzini}, Ronaldo and {Borotto}, Fabio and {Brez}, Alessandro and {Brienza}, Daniele and {Caporale}, Ciro and {Cardelli}, Claudia and {Carpentiero}, Rita and {Castellano}, Simone and {Castronuovo}, Marco and {Cavalli}, Luca and {Cavazzuti}, Elisabetta and {Ceccanti}, Marco and {Centrone}, Mauro and {Citraro}, Saverio and {Costa}, Enrico and {D'Alba}, Elisa and {D'Amico}, Fabio and {Del Monte}, Ettore and {Di Cosimo}, Sergio and {Di Lalla}, Niccol{\`o} and {Di Persio}, Giuseppe and {Donnarumma}, Immacolata and {Evangelista}, Yuri and {Ferrazzoli}, Riccardo and {Latronico}, Luca and {Lefevre}, Carlo and {Loffredo}, Pasqualino and {Lorenzi}, Paolo and {Lucchesi}, Leonardo and {Magazz{\`u}}, Carlo and {Magazz{\`u}}, Guido and {Maldera}, Simone and {Manfreda}, Alberto and {Mangraviti}, Elio and {Marengo}, Marco and {Matt}, Giorgio and {Mereu}, Paolo and {Minuti}, Massimo and {Morbidini}, Alfredo and {Mosti}, Federico and {Nasimi}, Hikmat and {Negri}, Barbara and {Nuti}, Alessio and {O'Dell}, Stephen L. and {Orsini}, Leonardo and {Perri}, Matteo and {Pesce-Rollins}, Melissa and {Piazzolla}, Raffaele and {Pieraccini}, Stefano and {Pilia}, Maura and {Pinchera}, Michele and {Profeti}, Alessandro and {Puccetti}, Simonetta and {Ramsey}, Brian D. and {Ratheesh}, Ajay and {Rubini}, Alda and {Santoli}, Francesco and {Sarra}, Paolo and {Scalise}, Emanuele and {Sciortino}, Andrea and {Sgr{\`o}}, Carmelo and {Spandre}, Gloria and {Tardiola}, Marcello and {Tennant}, Allyn F. and {Tobia}, Antonino and {Trois}, Alessio and {Vimercati}, Marco and {Weisskopf}, Martin C. and {Zanetti}, Davide and {Zanetti}, Francesco},
        title = "{Calibration of the IXPE Focal Plane X-Ray Polarimeters to Polarized Radiation}",
      journal = {\aj},
         year = 2022,
        month = sep,
       volume = {164},
       number = {3},
          eid = {103},
        pages = {103},
          doi = {10.3847/1538-3881/ac7719},
archivePrefix = {arXiv},
       eprint = {2206.07582},
 primaryClass = {astro-ph.IM},
       adsurl = {https://ui.adsabs.harvard.edu/abs/2022AJ....164..103D}
}

@ARTICLE{2022Doroshenko,
       author = {{Doroshenko}, Victor and {Poutanen}, Juri and {Tsygankov}, Sergey S. and {Suleimanov}, Valery F. and {Bachetti}, Matteo and {Caiazzo}, Ilaria and {Costa}, Enrico and {Di Marco}, Alessandro and {Heyl}, Jeremy and {La Monaca}, Fabio and {Muleri}, Fabio and {Mushtukov}, Alexander A. and {Pavlov}, George G. and {Ramsey}, Brian D. and {Rankin}, John and {Santangelo}, Andrea and {Soffitta}, Paolo and {Staubert}, R{\"u}diger and {Weisskopf}, Martin C. and {Zane}, Silvia and {Agudo}, Iv{\'a}n and {Antonelli}, Lucio A. and {Baldini}, Luca and {Baumgartner}, Wayne H. and {Bellazzini}, Ronaldo and {Bianchi}, Stefano and {Bongiorno}, Stephen D. and {Bonino}, Raffaella and {Brez}, Alessandro and {Bucciantini}, Niccol{\`o} and {Capitanio}, Fiamma and {Castellano}, Simone and {Cavazzuti}, Elisabetta and {Ciprini}, Stefano and {De Rosa}, Alessandra and {Del Monte}, Ettore and {Di Gesu}, Laura and {Di Lalla}, Niccol{\`o} and {Donnarumma}, Immacolata and {Dov{\v{c}}iak}, Michal and {Ehlert}, Steven R. and {Enoto}, Teruaki and {Evangelista}, Yuri and {Fabiani}, Sergio and {Ferrazzoli}, Riccardo and {Garcia}, Javier A. and {Gunji}, Shuichi and {Hayashida}, Kiyoshi and {Iwakiri}, Wataru and {Jorstad}, Svetlana G. and {Karas}, Vladimir and {Kitaguchi}, Takao and {Kolodziejczak}, Jeffery J. and {Krawczynski}, Henric and {Latronico}, Luca and {Liodakis}, Ioannis and {Maldera}, Simone and {Manfreda}, Alberto and {Marin}, Fr{\'e}d{\'e}ric and {Marinucci}, Andrea and {Marscher}, Alan P. and {Marshall}, Herman L. and {Matt}, Giorgio and {Mitsuishi}, Ikuyuki and {Mizuno}, Tsunefumi and {Ng}, Chi-Yung and {O'Dell}, Stephen L. and {Omodei}, Nicola and {Oppedisano}, Chiara and {Papitto}, Alessandro and {Peirson}, Abel L. and {Perri}, Matteo and {Pesce-Rollins}, Melissa and {Pilia}, Maura and {Possenti}, Andrea and {Puccetti}, Simonetta and {Ratheesh}, Ajay and {Romani}, Roger W. and {Sgr{\`o}}, Carmelo and {Slane}, Patrick and {Spandre}, Gloria and {Sunyaev}, Rashid A. and {Tamagawa}, Toru and {Tavecchio}, Fabrizio and {Taverna}, Roberto and {Tawara}, Yuzuru and {Tennant}, Allyn F. and {Thomas}, Nicolas E. and {Tombesi}, Francesco and {Trois}, Alessio and {Turolla}, Roberto and {Vink}, Jacco and {Wu}, Kinwah and {Xie}, Fei},
        title = "{Determination of X-ray pulsar geometry with IXPE polarimetry}",
      journal = {Nature Astronomy},
         year = 2022,
        month = dec,
       volume = {6},
        pages = {1433-1443},
          doi = {10.1038/s41550-022-01799-5},
archivePrefix = {arXiv},
       eprint = {2206.07138},
 primaryClass = {astro-ph.HE},
       adsurl = {https://ui.adsabs.harvard.edu/abs/2022NatAs...6.1433D}
}

@ARTICLE{2023Garg,
       author = {{Garg}, Akash and {Rawat}, Divya and {Bhargava}, Yash and {M{\'e}ndez}, Mariano and {Bhattacharyya}, Sudip},
        title = "{Flux-resolved Spectropolarimetric Evolution of the X-Ray Pulsar Hercules X-1 Using IXPE}",
      journal = {\apjl},
         year = 2023,
        month = may,
       volume = {948},
       number = {1},
          eid = {L10},
        pages = {L10},
          doi = {10.3847/2041-8213/acccfa},
archivePrefix = {arXiv},
       eprint = {2304.00889},
 primaryClass = {astro-ph.HE},
       adsurl = {https://ui.adsabs.harvard.edu/abs/2023ApJ...948L..10G}
}

@ARTICLE{2024Heyl,
       author = {{Heyl}, Jeremy and {Doroshenko}, Victor and {Gonz{\'a}lez-Caniulef}, Denis and {Caiazzo}, Ilaria and {Poutanen}, Juri and {Mushtukov}, Alexander and {Tsygankov}, Sergey S. and {Kirmizibayrak}, Demet and {Bachetti}, Matteo and {Pavlov}, George G. and {Forsblom}, Sofia V. and {Malacaria}, Christian and {Suleimanov}, Valery F. and {Agudo}, Iv{\'a}n and {Antonelli}, Lucio Angelo and {Baldini}, Luca and {Baumgartner}, Wayne H. and {Bellazzini}, Ronaldo and {Bianchi}, Stefano and {Bongiorno}, Stephen D. and {Bonino}, Raffaella and {Brez}, Alessandro and {Bucciantini}, Niccol{\`o} and {Capitanio}, Fiamma and {Castellano}, Simone and {Cavazzuti}, Elisabetta and {Chen}, Chien-Ting and {Ciprini}, Stefano and {Costa}, Enrico and {De Rosa}, Alessandra and {Del Monte}, Ettore and {Di Gesu}, Laura and {Di Lalla}, Niccol{\`o} and {Di Marco}, Alessandro and {Donnarumma}, Immacolata and {Dov{\v{c}}iak}, Michal and {Ehlert}, Steven R. and {Enoto}, Teruaki and {Evangelista}, Yuri and {Fabiani}, Sergio and {Ferrazzoli}, Riccardo and {Garcia}, Javier A. and {Gunji}, Shuichi and {Hayashida}, Kiyoshi and {Iwakiri}, Wataru and {Jorstad}, Svetlana G. and {Kaaret}, Philip and {Karas}, Vladimir and {Kislat}, Fabian and {Kitaguchi}, Takao and {Kolodziejczak}, Jeffery J. and {Krawczynski}, Henric and {La Monaca}, Fabio and {Latronico}, Luca and {Liodakis}, Ioannis and {Maldera}, Simone and {Manfreda}, Alberto and {Marin}, Fr{\'e}d{\'e}ric and {Marinucci}, Andrea and {Marscher}, Alan P. and {Marshall}, Herman L. and {Massaro}, Francesco and {Matt}, Giorgio and {Mitsuishi}, Ikuyuki and {Mizuno}, Tsunefumi and {Muleri}, Fabio and {Negro}, Michela and {Ng}, C.-Y. and {O'Dell}, Stephen L. and {Omodei}, Nicola and {Oppedisano}, Chiara and {Papitto}, Alessandro and {Peirson}, Abel Lawrence and {Perri}, Matteo and {Pesce-Rollins}, Melissa and {Petrucci}, Pierre-Olivier and {Pilia}, Maura and {Possenti}, Andrea and {Puccetti}, Simonetta and {Ramsey}, Brian D. and {Rankin}, John and {Ratheesh}, Ajay and {Roberts}, Oliver J. and {Romani}, Roger W. and {Sgr{\`o}}, Carmelo and {Slane}, Patrick and {Soffitta}, Paolo and {Spandre}, Gloria and {Swartz}, Douglas A. and {Tamagawa}, Toru and {Tavecchio}, Fabrizio and {Taverna}, Roberto and {Tawara}, Yuzuru and {Tennant}, Allyn F. and {Thomas}, Nicholas E. and {Tombesi}, Francesco and {Trois}, Alessio and {Turolla}, Roberto and {Vink}, Jacco and {Weisskopf}, Martin C. and {Wu}, Kinwah and {Xie}, Fei and {Zane}, Silvia},
        title = "{Complex rotational dynamics of the neutron star in Hercules X-1 revealed by X-ray polarization}",
      journal = {Nature Astronomy},
         year = 2024,
        month = aug,
       volume = {8},
        pages = {1047-1053},
          doi = {10.1038/s41550-024-02295-8},
       adsurl = {https://ui.adsabs.harvard.edu/abs/2024NatAs...8.1047H}
}

@ARTICLE{2024Zhao,
       author = {{Zhao}, Q.~C. and {Li}, H.~C. and {Tao}, L. and {Feng}, H. and {Zhang}, S.~N. and {Walter}, R. and {Ge}, M.~Y. and {Tong}, H. and {Ji}, L. and {Zhang}, L. and {Qu}, J.~L. and {Huang}, Y. and {Ma}, X. and {Zhang}, S. and {Yin}, Q.~Q. and {Yin}, H.~X. and {Ma}, R.~C. and {Zhao}, S.~J. and {Li}, P.~P. and {Yang}, Z.~X. and {Liu}, H.~X. and {Yu}, W. and {Huang}, Y.~M. and {Li}, Z.~X. and {Li}, Y.~J. and {Xiao}, J.~Y. and {Zhao}, K.},
        title = "{Polarization perspectives on Hercules X-1: further constraining the geometry}",
      journal = {\mnras},
         year = 2024,
        month = jul,
       volume = {531},
       number = {4},
        pages = {3935-3949},
          doi = {10.1093/mnras/stae1173},
archivePrefix = {arXiv},
       eprint = {2405.00509},
 primaryClass = {astro-ph.HE},
       adsurl = {https://ui.adsabs.harvard.edu/abs/2024MNRAS.531.3935Z}
}

@ARTICLE{2023Cocchi,
       author = {{Cocchi}, Massimo and {Gnarini}, Andrea and {Fabiani}, Sergio and {Ursini}, Francesco and {Poutanen}, Juri and {Capitanio}, Fiamma and {Bobrikova}, Anna and {Farinelli}, Ruben and {Paizis}, Adamantia and {Sidoli}, Lara and {Veledina}, Alexandra and {Bianchi}, Stefano and {Di Marco}, Alessandro and {Ingram}, Adam and {Kajava}, Jari J.~E. and {La Monaca}, Fabio and {Matt}, Giorgio and {Malacaria}, Christian and {Miku{\v{s}}incov{\'a}}, Romana and {Rankin}, John and {Zane}, Silvia and {Agudo}, Iv{\'a}n and {Antonelli}, Lucio A. and {Bachetti}, Matteo and {Baldini}, Luca and {Baumgartner}, Wayne H. and {Bellazzini}, Ronaldo and {Bongiorno}, Stephen D. and {Bonino}, Raffaella and {Brez}, Alessandro and {Bucciantini}, Niccol{\`o} and {Castellano}, Simone and {Cavazzuti}, Elisabetta and {Chen}, Chien-Ting and {Ciprini}, Stefano and {Costa}, Enrico and {De Rosa}, Alessandra and {Del Monte}, Ettore and {Di Gesu}, Laura and {Di Lalla}, Niccol{\`o} and {Donnarumma}, Immacolata and {Doroshenko}, Victor and {Dov{\v{c}}iak}, Michal and {Ehlert}, Steven R. and {Enoto}, Teruaki and {Evangelista}, Yuri and {Ferrazzoli}, Riccardo and {Garcia}, Javier A. and {Gunji}, Shuichi and {Hayashida}, Kiyoshi and {Heyl}, Jeremy and {Iwakiri}, Wataru and {Jorstad}, Svetlana G. and {Kaaret}, Philip and {Karas}, Vladimir and {Kislat}, Fabian and {Kitaguchi}, Takao and {Kolodziejczak}, Jeffery J. and {Krawczynski}, Henric and {Latronico}, Luca and {Liodakis}, Ioannis and {Maldera}, Simone and {Manfreda}, Alberto and {Marin}, Fr{\'e}d{\'e}ric and {Marinucci}, Andrea and {Marscher}, Alan P. and {Marshall}, Herman L. and {Massaro}, Francesco and {Mitsuishi}, Ikuyuki and {Mizuno}, Tsunefumi and {Muleri}, Fabio and {Negro}, Michela and {Ng}, Chi-Yung and {O'Dell}, Stephen L. and {Omodei}, Nicola and {Oppedisano}, Chiara and {Papitto}, Alessandro and {Pavlov}, George G. and {Peirson}, Abel L. and {Perri}, Matteo and {Pesce-Rollins}, Melissa and {Petrucci}, Pierre-Olivier and {Pilia}, Maura and {Possenti}, Andrea and {Puccetti}, Simonetta and {Ramsey}, Brian D. and {Ratheesh}, Ajay and {Roberts}, Oliver J. and {Romani}, Roger W. and {Sgr{\`o}}, Carmelo and {Slane}, Patrick and {Soffitta}, Paolo and {Spandre}, Gloria and {Swartz}, Douglas A. and {Tamagawa}, Toru and {Tavecchio}, Fabrizio and {Taverna}, Roberto and {Tawara}, Yuzuru and {Tennant}, Allyn F. and {Thomas}, Nicholas E. and {Tombesi}, Francesco and {Trois}, Alessio and {Tsygankov}, Sergey S. and {Turolla}, Roberto and {Vink}, Jacco and {Weisskopf}, Martin C. and {Wu}, Kinwah and {Xie}, Fei},
        title = "{Discovery of strongly variable X-ray polarization in the neutron star low-mass X-ray binary transient XTE J1701‒462}",
      journal = {\aap},
         year = 2023,
        month = jun,
       volume = {674},
          eid = {L10},
        pages = {L10},
          doi = {10.1051/0004-6361/202346275},
archivePrefix = {arXiv},
       eprint = {2306.10965},
 primaryClass = {astro-ph.HE},
       adsurl = {https://ui.adsabs.harvard.edu/abs/2023A&A...674L..10C}
}

@ARTICLE{2022Baldini,
       author = {{Baldini}, Luca and {Bucciantini}, Niccol{\`o} and {Lalla}, Niccol{\`o} Di and {Ehlert}, Steven and {Manfreda}, Alberto and {Negro}, Michela and {Omodei}, Nicola and {Pesce-Rollins}, Melissa and {Sgr{\`o}}, Carmelo and {Silvestri}, Stefano},
        title = "{ixpeobssim: A simulation and analysis framework for the imaging X-ray polarimetry explorer}",
      journal = {SoftwareX},
         year = 2022,
        month = jul,
       volume = {19},
          eid = {101194},
        pages = {101194},
          doi = {10.1016/j.softx.2022.101194},
archivePrefix = {arXiv},
       eprint = {2203.06384},
 primaryClass = {astro-ph.IM},
       adsurl = {https://ui.adsabs.harvard.edu/abs/2022SoftX..1901194B}
}

@ARTICLE{2023DiMarco,
       author = {{Di Marco}, Alessandro and {Soffitta}, Paolo and {Costa}, Enrico and {Ferrazzoli}, Riccardo and {La Monaca}, Fabio and {Rankin}, John and {Ratheesh}, Ajay and {Xie}, Fei and {Baldini}, Luca and {Del Monte}, Ettore and {Ehlert}, Steven R. and {Fabiani}, Sergio and {Kim}, Dawoon E. and {Muleri}, Fabio and {O'Dell}, Stephen L. and {Ramsey}, Brian D. and {Rubini}, Alda and {Sgr{\`o}}, Carmelo and {Silvestri}, Stefano and {Tennant}, Allyn F. and {Weisskopf}, Martin C.},
        title = "{Handling the Background in IXPE Polarimetric Data}",
      journal = {\aj},
         year = 2023,
        month = apr,
       volume = {165},
       number = {4},
          eid = {143},
        pages = {143},
          doi = {10.3847/1538-3881/acba0f},
archivePrefix = {arXiv},
       eprint = {2302.02927},
 primaryClass = {astro-ph.IM},
       adsurl = {https://ui.adsabs.harvard.edu/abs/2023AJ....165..143D}
}

@INPROCEEDINGS{2022a_DiMarco,
       author = {{Di Marco}, Alessandro and {Tennant}, Allyn and {La Monaca}, Fabio and {Muleri}, Fabio and {Rankin}, John and {Rushing}, John and {Soffitta}, Paolo and {Baglioni}, Giancarlo and {Baldini}, Luca and {Costa}, Enrico and {Dietz}, Kurt and {Fabiani}, Sergio and {Latorre}, Vittorio and {Locatelli}, Ugo and {Manfreda}, Alberto and {O'Dell}, Stephen and {Peirson}, Lawrence and {Weisskopf}, Martin},
        title = "{Validation of neural network software by using IXPE ground calibration data}",
    booktitle = {Space Telescopes and Instrumentation 2022: Ultraviolet to Gamma Ray},
         year = 2022,
       editor = {{den Herder}, Jan-Willem A. and {Nikzad}, Shouleh and {Nakazawa}, Kazuhiro},
       series = {Society of Photo-Optical Instrumentation Engineers (SPIE) Conference Series},
       volume = {12181},
        month = aug,
          eid = {1218157},
        pages = {1218157},
          doi = {10.1117/12.2628976},
       adsurl = {https://ui.adsabs.harvard.edu/abs/2022SPIE12181E..57D}
}

@ARTICLE{2021Peirson,
       author = {{Peirson}, A.~L. and {Romani}, Roger W.},
        title = "{Toward Optimal Signal Extraction for Imaging X-Ray Polarimetry}",
      journal = {\apj},
         year = 2021,
        month = oct,
       volume = {920},
       number = {1},
          eid = {40},
        pages = {40},
          doi = {10.3847/1538-4357/ac157d},
archivePrefix = {arXiv},
       eprint = {2107.08289},
 primaryClass = {astro-ph.IM},
       adsurl = {https://ui.adsabs.harvard.edu/abs/2021ApJ...920...40P}
}

@misc{latorre2026,
      title={Graph Neural Networks Trained on Null Signal for Angle Reconstruction in X-ray Polarimetry}, 
      author={Vittorio Latorre and Victor Rodriguez-Fernandez and Alessandro Di Marco and Fabio La Monaca and Fabio Muleri and Paolo Soffitta},
      year={2026},
      eprint={2606.16922},
      archivePrefix={arXiv},
      primaryClass={astro-ph.IM},
      url={https://arxiv.org/abs/2606.16922}, 
}

@software{container,
  author       = {Silvestri, Stefano and
                  Cibrario, Nicolò and
                  Negro, Michela and
                  Moriakov, Nikita},
  title        = {Hybrid reconstruction docker for IXPE data},
  month        = jul,
  year         = 2026,
  publisher    = {Zenodo},
  doi          = {10.5281/zenodo.21395803},
  url          = {https://doi.org/10.5281/zenodo.21395803},
}

@INPROCEEDINGS{2010Weisskopf,
       author = {{Weisskopf}, Martin C. and {Elsner}, Ronald F. and {O'Dell}, Stephen L.},
        title = "{On understanding the figures of merit for detection and measurement of x-ray polarization}",
    booktitle = {Space Telescopes and Instrumentation 2010: Ultraviolet to Gamma Ray},
         year = 2010,
       editor = {{Arnaud}, Monique and {Murray}, Stephen S. and {Takahashi}, Tadayuki},
       series = {Society of Photo-Optical Instrumentation Engineers (SPIE) Conference Series},
       volume = {7732},
        month = jul,
          eid = {77320E},
        pages = {77320E},
          doi = {10.1117/12.857357},
archivePrefix = {arXiv},
       eprint = {1006.3711},
 primaryClass = {astro-ph.IM},
       adsurl = {https://ui.adsabs.harvard.edu/abs/2010SPIE.7732E..0EW}
}

@article{1936Bonferroni,
  author   = {Bonferroni, Carlo E.},
  title    = {Teoria statistica delle classi e calcolo delle probabilit\`a},
  journal  = {Pubblicazioni del R. Istituto Superiore di Scienze Economiche e Commerciali di Firenze},
  volume   = {8},
  pages    = {3--62},
  year     = {1936}
}
\bibliographystyle{aa}
\end{document}